\documentstyle[12pt,epsf]{article}
\def\TC{T_c}                  \def\tc{\ifmmode\TC\else$\TC$\fi}

\def\conlim{\mathop{\longrightarrow}\limits_{\scriptscriptstyle T\to\infty}}
\def\spose#1{\hbox to 0pt{#1\hss}}
\def\ltapprox{\mathrel{\spose{\lower 3pt\hbox{$\mathchar"218$}}
 \raise 2.0pt\hbox{$\mathchar"13C$}}}
\def\gtapprox{\mathrel{\spose{\lower 3pt\hbox{$\mathchar"218$}}
 \raise 2.0pt\hbox{$\mathchar"13E$}}}
\def\inapprox{\mathrel{\spose{\lower 3pt\hbox{$\mathchar"218$}}
 \raise 2.0pt\hbox{$\mathchar"232$}}}
\def\eg{{\sl e.g.\/}}

\def\ie{{\sl i.e.\/}}
\newcommand{\beq}{\begin{equation}} \newcommand{\eeq}{\end{equation}}
\newcommand{\beqa}{\begin{eqnarray}}\newcommand{\eeqa}{\end{eqnarray}}
 \newcommand{\Tr}{\mbox{Tr}}

\begin{document}
\thispagestyle{empty}
\mbox{} \hspace{11.76truecm} FSU-SCRI-94-71\\
\mbox{} \hspace{12.50truecm} HU Berlin-IEP-94/10\\
\mbox{} \hspace{12.40truecm} July 15, 1994\\

\begin{center}
\vspace*{1.0cm}
{\large SCALING IN THE POSITIVE PLAQUETTE MODEL AND\\
        UNIVERSALITY IN SU(2) LATTICE GAUGE THEORY}\\
\vspace*{1.0cm}
{\large J.~Fingberg,$^a$ U.M.~Heller$^a$ and V.~Mitrjushkin$^b$}
\\
\vspace*{1.0cm}
{\normalsize 
$\mbox{}^a$ {SCRI, Florida State University, Tallahassee, FL 32306-4052, USA.}\\
$\mbox{}^b$ {Institut f\"ur Physik,
Humboldt--Universit\"at zu Berlin, 10099 Berlin, Germany.}
\footnote{Permanent address: Joint Institute for Nuclear Research, 
Dubna, Russia}
}\\

\vspace*{2cm}
{\large \bf Abstract}
\end{center}

\setlength{\baselineskip}{1.3\baselineskip}
We investigate universality, scaling, the $\beta$--function and the 
topological charge in the positive plaquette model for SU(2) lattice
gauge theory. Comparing physical quantities, like the critical temperature,
the string tension, glueball masses, and their ratios, we explore the effect
of a complete suppression of a certain lattice artifact, namely the negative
plaquettes, for SU(2) lattice gauge theory. Our result is that this
modification does 
not change the continuum limit, \ie, the universality class.
The positive plaquette model and the standard Wilson formulation describe the
same physical situation. The approach to the continuum limit given by the
$\beta$--function in terms of the bare lattice coupling, however, is rather
different: the  $\beta$--function of the positive plaquette model does not
show a ``dip'' like the model with standard Wilson action.

\newpage\setcounter{page}1
\section{Introduction}\label{sec:introduction}

The pure gauge sector of lattice QCD, though at present the most 
studied part of the theory in numerical simulations,
still provides many open questions.
One of the most interesting yet unresolved problems is 
connected with the mechanism of confinement at low temperatures.
Among the possible solutions to this problem one could mention
the dual superconductivity mechanism \cite{thooft} or the
effective $Z_2$ theory of confinement \cite{mack}
based on a $Z_2$--vortex  condensation mechanism \cite{mack_p}.
However, as is well--known, the lattice formulation of
any quantum field theory is not unique. Different
lattice theories, equivalent on a classical level,
may possess completely different phase structures. 
As a result these models can have different continuum limits,
\ie, belong to different universality classes
(for a recent example see~\cite{Sokal}).
In order to get reliable ``physical'' results for continuum
physics from numerical calculations, it is therefore necessary to compare
different forms of lattice actions.

The question of the choice of one or another lattice
formulation can also be connected to the problem of 
lattice artifacts (for a short review see \eg ~\cite{B_C_M} 
and references therein).
For example, small--scale ($\sim a$) fluctuations carrying a
nontrivial topological charge can lead to the divergence
of the topological susceptibility in the continuum limit
(at least, for the geometric definition of the topological charge)
\cite{pt,desy}. A source of trouble in this case are the special 
field configurations containing negative plaquettes,
$\frac{1}{2} \Tr U_{P} \sim -1$ (for $SU(2)$ gauge group).
In more general form the question could be formulated as:
how do {\it non--smooth} (rough) lattice field configurations
influence the calculation of physical observables?

Another problem of general interest includes bad scaling
behavior of SU(N) gauge theories.
Previous studies with the standard Wilson plaquette action 
(SWA) on lattices up to
$48^3 \times 16$ at finite temperature~\cite{scaling} and up to
$48^3 \times 56$ at zero temperature~\cite{UKQCD_92}
showed that (asymptotic) scaling violating terms are still strong
at the $\beta$--values accessible to todays numerical simulations.
These terms manifest themselves most strikingly in the famous dip
of the ``step $\beta$--function'', $\Delta \beta(\beta)$.
This pattern is similar in all SU(N) gauge theories with a wide set
of different actions.

A lot of promising attempts to overcome these difficulties are based on
the use of ``improved'' actions and/or effective coupling schemes.
The latter, of course, only affect the asymptotic scaling behavior, now
in terms of the effective, instead of the bare, coupling. Improved
actions, on the other hand, should diminish scaling violations and
improve the control over dislocations. At the same time they should be
in the same universality class as the standard Wilson action, and hence
lead to the same continuum limit.

This work is devoted to the study of the long--distance properties
in the theory with a modified action where negative plaquettes are
suppressed (positive plaquette model).
Previous investigations of the positive plaquette model (PPM), first
considered by Mack and Pietarinen~\cite{M_P} and 
Bornyakov, Creutz and Mitrjushkin~\cite{B_C_M} (see also ~\cite{at}),
suggest that large--scale objects, \eg ~Polyakov loop correlators,
Wilson loops and Creutz ratios, are strongly influenced by 
lattice artifacts even at comparatively large values of the 
inverse coupling $\beta$. 
The size--dependence of Wilson loops was shown to be consistent
with an area law but the scaling properties of the string
tension, \ie, the fate of confinement in this model,
were left beyond the scope of both papers~\cite{M_P,B_C_M}.

The dip in the ``step $\beta$--function'', $\Delta\beta(\beta)$,
observed for the standard model might be associated with the presence of
negative plaquettes, definite lattice artifacts which vanishe in
the continuum limit.
This dip has been associated with the phase structure when adding
an adjoint coupling, insensitive to the sign of $~\Tr U_p~$, to the
standard fundamental one. The phase structure stems from the fact
that as $\beta_a \to \infty$ the plaquettes are frozen to $\pm 1$
\cite{Bhanot_Creutz}, but in the PPM the value $-1$ is not allowed
and therefore the phase transition at large $\beta_a$, whose critical
endpoint close to the fundamental axis is supposed to be responsible
for the dip, does not occur.
The interesting question then is, whether the removal of
negative plaquettes really avoids this dip and whether it
improves the approach to the continuum limit.

A possible verification of this scenario requires a test of universality
and a comparison of the scaling and asymptotic scaling behavior in the PPM 
and the conventional SU(2) lattice gauge theory.
This task has been accomplished by
a comprehensive study of the PPM both at finite and zero temperature.
Our strategy consisted of two steps. First we determined the critical
couplings for various lattice sizes and measured some universal quantities.
These results were then used in an investigation of the model at vanishing
temperature which consisted of a measurement of the heavy quark potential and
the glueball spectrum of the two lowest excitations.
We found it important to study a wide set of mass ratios in order to test
whether some of these operators are influenced more strongly than others.
Our work was completed by an MCRG analysis of the step $\beta$--function and
a measurement of the topological charge and susceptibility.
It should be noted that the removal of negative plaquettes is an
entirely non--perturbative modification of the Wilson action. The
perturbation theory remains unchanged. In particular, the
$\Lambda$--parameter remains unchanged, as does the inverse coupling
in the continuum limit, \ie, $\beta_{PPM} = \beta_W$ as $\beta_W \to
\infty$.

For the present study we have chosen a complete suppression of negative
plaquettes, \ie, a cutoff for the plaquettes at zero. Any other value
smaller than $1$ would have been possible for this cutoff. However, as
we will see, the critical coupling $\beta_c$ for the deconfinement
transition in the PPM for $N_\tau=2$ is already very small. Therefore,
for choices of the cutoff larger than zero, lattices with a small extent
in the time direction might no longer be confined at positive values of
the gauge coupling. While this does not have to affect the continuum
limit, it makes a numerical investigation far more difficult, since much
larger lattices would need to be considered.

In the following section we describe the details of our simulations.
Then we discuss our results obtained at finite temperature. In sections
\ref{sec:potential} and \ref{sec:glueball} we present our results for
the heavy quark potential and the glueball spectrum. 
The next section is dedicated to a detailed MCRG analysis of the step
$\beta$--function. In section \ref{sec:scaling} we combine these findings and
discuss their implication for scaling and asymptotic scaling.
Finally we summarize our results and come to the conclusions.

\section{The simulations}\label{sec:simulation}

In this work the calculations were done on symmetric lattices 
with size up to $16^4$ and on asymmetric ones with
temporal extent $N_{\tau} = 2, \, 4$ and $8$ and spatial extent up to
$N_\sigma = 24$.
The  values of the gauge coupling  $\beta$ we used covered
the interval between zero and $2.6$.

For most of the simulations, whose results will be described in
the subsequent sections, we have used a mixture of 4 microcanonical
overrelaxation sweeps followed by one heat bath sweep. In the
microcanonical step
\beq U_l \to U_l^\prime = W^\dagger U_l^\dagger
W^\dagger
\eeq 
where $W$ is the ``staple'' projected onto the SU(2) group, one or more of
the 6 plaquettes which contain the link $l$ can become negative. If this
occurs, the change is, of course, rejected. This rejection rate, quite large
at small $\beta$, fortunately decreases rapidly with increasing $\beta$
where the autocorrelation time becomes larger and the overrelaxation steps
therefore more important.

The heat bath updating was done for smaller $\beta$ using the Creutz
algorithm \cite{Cr_HB} and for larger $\beta$ the Kennedy--Pendleton version
\cite{KP_HB}. The choice of version was done by measuring the CPU time of
short trial runs. In both heat bath versions we added, of course, rejection
of trial link updates that produce negative plaquettes to the usual
criterion.

In the runs at very small $\beta$'s the heat bath acceptance rate dropped
dramatically due to the frequency of proposed links that give negative
plaquettes and have to be rejected. There we used, instead, a Metropolis
algorithm with the size of the proposed random changes tuned to give an
acceptance rate of about 50\%.

We performed simulations on asymmetric lattices with $N_\tau = 2$, 4 and 8 to
study the deconfinement transition. In those simulations we measured the
Polyakov line after every heat bath (Metropolis) sweep. In the simulations
on symmetric lattices, used for the MCRG study and the measurement of the
string tension, glueball masses and topological susceptibility, we
typically measured after every 20 heat bath sweeps. In this way, about the
same amount of CPU time was spent on updating and on all the measurements.
It also turned out that this spacing --- recall that 4 overrelaxation sweeps
were done before every heat bath sweep --- made the measurements essentially
independent, with integrated autocorrelation length $\tau_{auto}\ltapprox 1$.

\section{The deconfinement temperature}\label{sec:Tc}

To check the universality of continuum
physics obtained with the PPM and the theory with the SWA, 
one should compare results obtained in ``equal physical circumstances''. 
In particular the lattice coupling must be chosen carefully. The easiest way
to do this is to study the deconfinement transition on lattices with the
same temporal extent $N_\tau$. This will ensure that we compare the two
theories at equal lattice spacings. Later on we will determine other
physical (zero--temperature) observables, like the string tension and
glueball masses, at the critical couplings. This will allow us a direct
comparison of ratios like $T_c/\sqrt{\sigma}$ at equal lattice spacings in
physical units.

The PPM, like the SU(2) theory with SWA, has a global $Z_2$ symmetry, which
gets spontaneously broken when the system crosses the critical temperature.
The transition is therefore expected to be of second order, in the
universality class of the 3 dimensional Ising model. However some care is
necessary, since universality may get disrupted by a change of the action
and therefore should be submitted to tests.
(It is worthwhile to note that there is no finite temperature phase
transition at $N_{\tau}=1$ at least at positive $\beta$'s~\cite{B_C_M}).

Finite--size scaling (FSS) techniques have proven to be effective tools to
explore the critical properties of statistical systems \cite{FSS}. In
combination with the density of states method a comprehensive study of the
region of the deconfinement transition becomes possible. We investigated
the PPM at finite temperature by measuring the order parameter, which is the
expectation value of the Polyakov loop,
$\langle |L|\rangle$, its susceptibility $\chi$
and its $4^{\rm th}$--order (Binder) cumulant, $g_4$,
\cite{Bin_cum} in order to determine critical couplings and test
universality by a consistency check of critical exponents. A completely
independent determination of the critical exponents would have
been beyond the scope of this paper.

\begin{figure}[htb]
\begin{center}
\vskip -1.0truecm
\leavevmode
\hbox{
\epsfysize=250pt\epsfbox{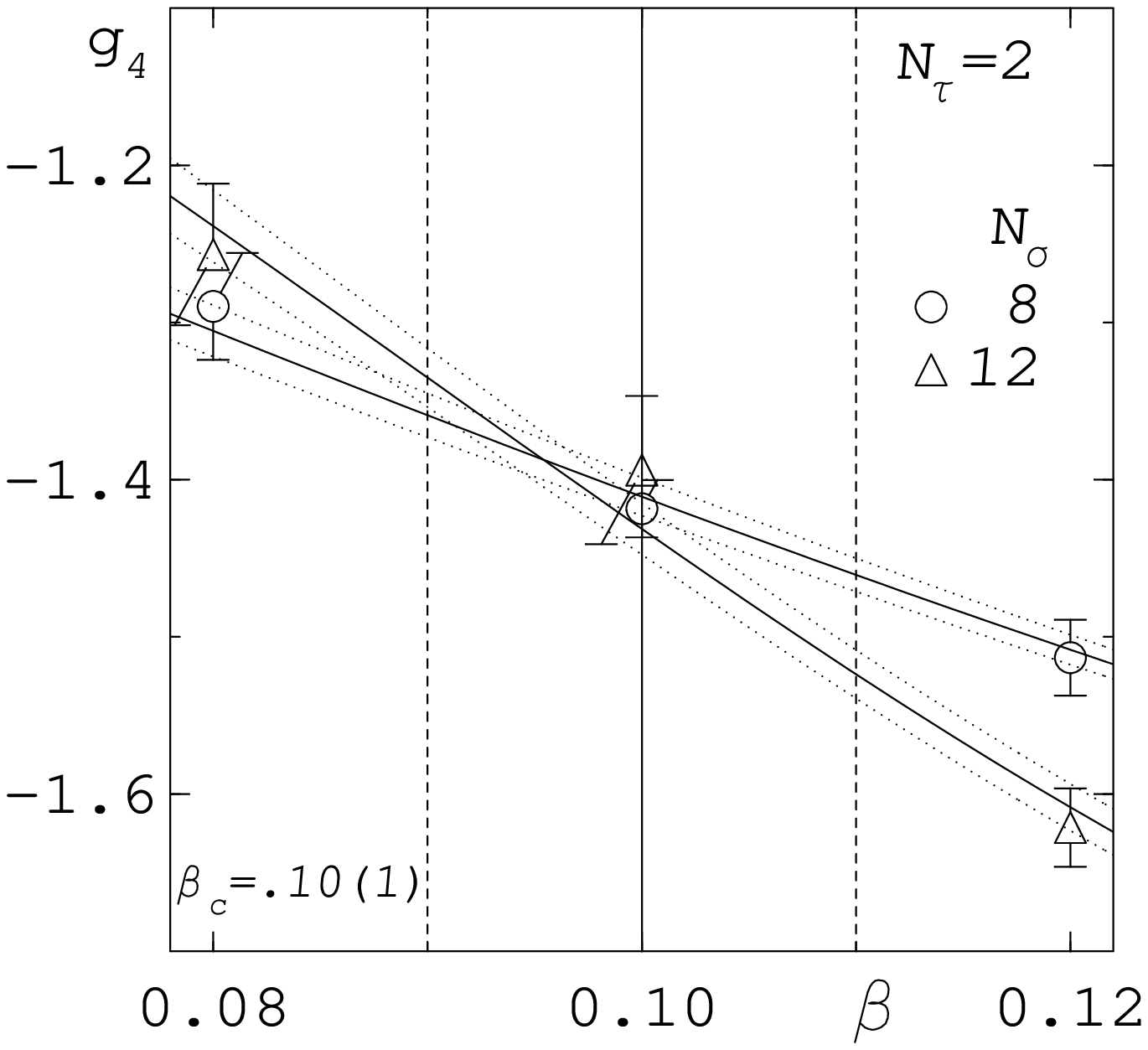} \hskip -1.0truecm
\epsfysize=250pt\epsfbox{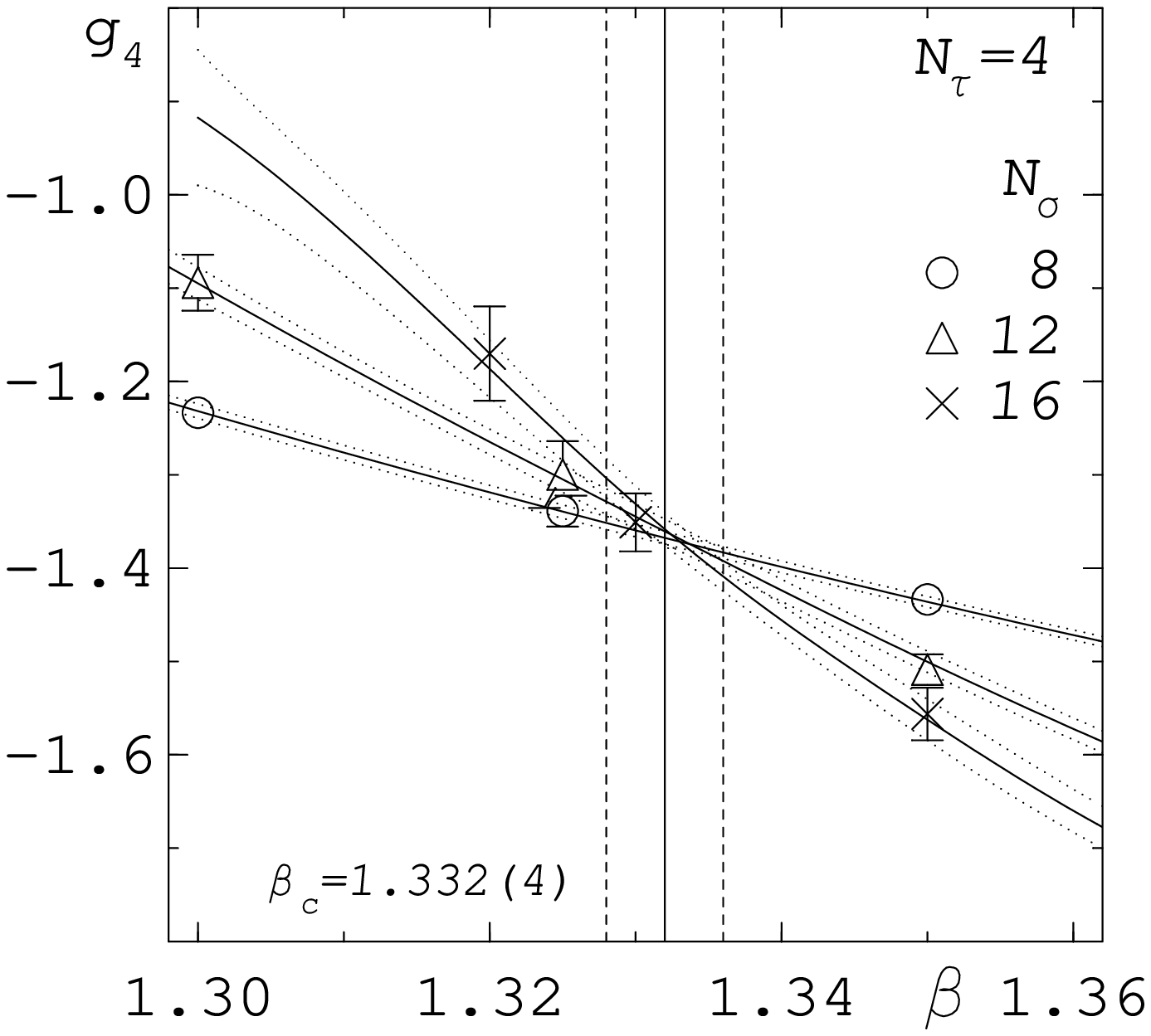} \hskip -1.0truecm
\epsfysize=250pt\epsfbox{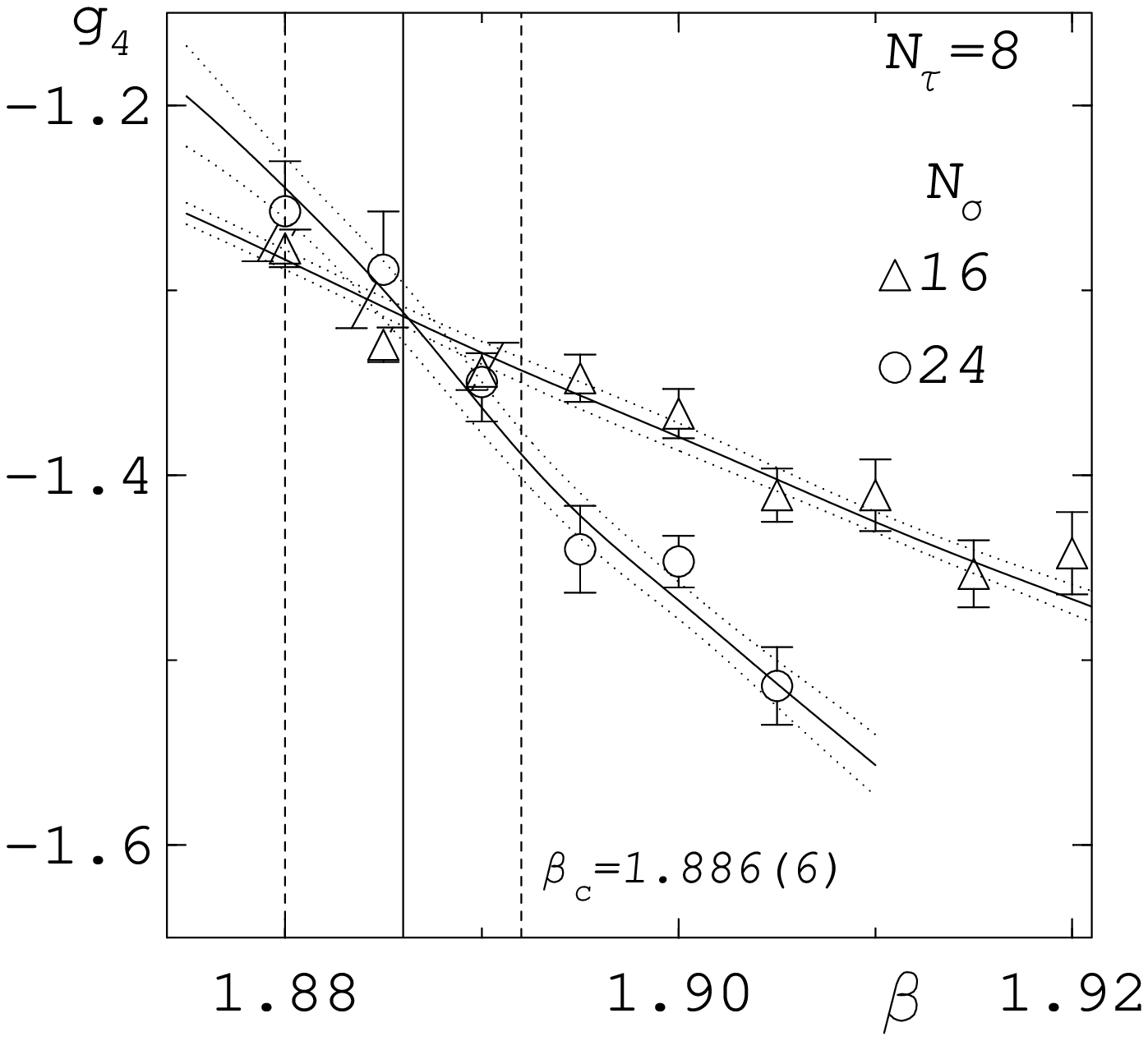}
     }
\end{center}
\vskip -2.5truecm
\caption{The $4^{\rm th}$ order cumulant $g_4$ for $N_\tau$=2, 4 and 8.
         The continous curves together with their error bands were obtained
         by standard reweighting.}
\label{fig:g4}
\end{figure}

We investigated the phase transition for systems with temporal extent
$N_\tau =2$, 4 and 8. The most extensive study was done for $N_\tau =4$,
where we used three different spatial lattice sizes, while for the others
we only considered two. The parameter values for our finite temperature
runs as well as the results for the Polyakov loop expectation values  are
summarized in Tab.~\ref{tab:FT}. We determined the critical couplings from
the crossing of curves of $g_4$ versus $\beta$
belonging to different $N_\sigma$ at fixed $N_\tau$.
The result is shown in Fig.~\ref{fig:g4}. The values we extracted for the
critical couplings are listed in Tab.~\ref{tab:betac}. A comparison with
results for the SWA shows a huge shift in the critical coupling, when
compared at fixed value of $N_\tau=2,~4$ and $8$. 
This large shift at strong coupling should be contrasted to the fact
that in the continuum limit, \ie, for very large $N_\tau$, the critical
couplings should become identical. Thus we can already conclude that
the ``step $\beta$--functions'' for the PPM and with SWA will be quite
different.

The value of the cumulant $g_4$ at the critical couplings is expected to be
a universal quantity. In all cases, $N_\tau=2,~4$ and $8$, it is in good
agreement with the value $g_4=-1.38(5)$, which was found for the
SWA~\cite{pheno}.
We find that the Binder cumulant $g_4$ shows the same FSS
behavior for the PPM, that already has been verified for the
SWA~\cite{pheno,scaling}.
The low temperature phase susceptibility $\chi_v$ is
expected~\cite{fss_su2,efficient} to behave like
\beq
\chi_v = N_\sigma^d~\langle L^2\rangle
       = (N_\sigma /N_\tau)^\frac{\gamma}{\nu}~~Q_{\chi_v}(g_4;N_\tau)
\label{eq:chi}
\eeq

\begin{figure}[htb]
\begin{center}
\vskip -1.0truecm
\leavevmode
\epsfysize=380pt\epsfbox{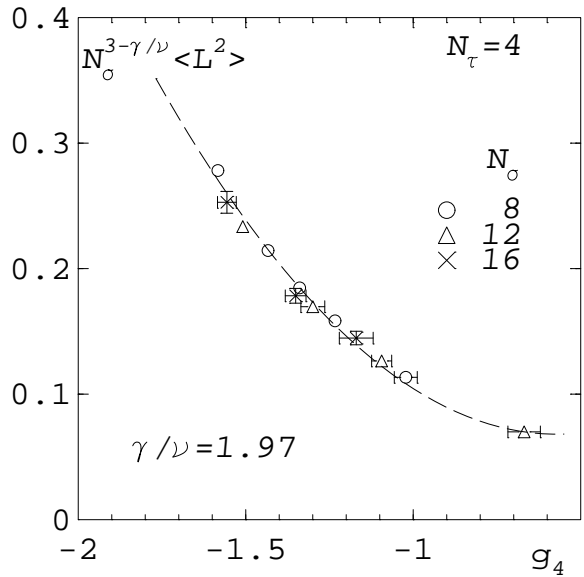}
\end{center}
\vskip -3.6truecm
\caption{Finite--size scaling of the susceptibility.
         The curve is a parabola drawn to guide the eye.}
\label{fig:scl_l2_t4}
\end{figure}

A FSS test~\cite{efficient} for $N_\tau =4$, where we have 3 different
lattice sizes, based on Eq.~(\ref{eq:chi}) consists of plotting the
scaling function $\tilde Q_{\chi_v}={N_\sigma}^{d-\frac{\gamma}{\nu}}~<L^2>$
as a function of $g_4$. Inserting the value $\gamma/\nu=1.970(11)$ of the
3--dimensional Ising model \cite{FerrLand} this test has the advantage that
there are no free parameters left. The result is shown in
Fig.~\ref{fig:scl_l2_t4}. We expect that data points coming from different
volumes fall on a unique curve, if finite--size scaling with the Ising
value of the ratio $\gamma /\nu$ holds. Our data points shown in
Fig.~\ref{fig:scl_l2_t4} agree very well with this assumption.

\begin{table}[htb]
\begin{center}
\begin{tabular}{|r|c|c|c|c|c|c|} \hline
$N_\sigma$&$N_\tau$&$\beta$ & meas.&$<|L|>$ & $<L^2>$&$g_4$     \\ \hline\hline
 8 & 2&0.080 &40000 & 0.1695(29) & 0.3675( 88)~$10^{-1}$ & -1.290(34) \\ \hline
 ~ & ~&0.100 &40000 & 0.1868(25) & 0.4300( 90)~$10^{-1}$ & -1.418(18) \\ \hline
 ~ & ~&0.120 &40000 & 0.2044(35) & 0.4985(131)~$10^{-1}$ & -1.513(24) \\ \hline
12 & 2&0.080 &40000 & 0.1326(32) & 0.2252( 83)~$10^{-1}$ & -1.257(45) \\ \hline
 ~ & ~&0.100 &40000 & 0.1536(42) & 0.2921(113)~$10^{-1}$ & -1.394(47) \\ \hline
 ~ & ~&0.120 &40000 & 0.1843(36) & 0.3904(119)~$10^{-1}$ & -1.621(25) \\ \hline
 8 & 4&1.250 &20000 & 0.0991(16) & 1.3317(330)~$10^{-2}$ & -1.022(34) \\ \hline
 ~ & ~&1.300 &80000 & 0.1200(10) & 1.8602(246)~$10^{-2}$ & -1.233(16) \\ \hline
 ~ & ~&1.325 &80000 & 0.1315(12) & 2.1686(284)~$10^{-2}$ & -1.339(17) \\ \hline
 ~ & ~&1.350 &60000 & 0.1436( 6) & 2.5180(154)~$10^{-2}$ & -1.434( 7) \\ \hline
 ~ & ~&1.400 &20000 & 0.1675(13) & 3.2667(392)~$10^{-2}$ & -1.583(12) \\ \hline
12 & 4&1.250 &20000 & 0.0608(16) & 0.5411(232)~$10^{-2}$ & -0.669(48) \\ \hline
 ~ & ~&1.300 &60000 & 0.0855(13) & 0.9782(221)~$10^{-2}$ & -1.094(30) \\ \hline
 ~ & ~&1.325 &60000 & 0.1017(19) & 1.3126(364)~$10^{-2}$ & -1.300(36) \\ \hline
 ~ & ~&1.350 &50000 & 0.1229(13) & 1.8061(283)~$10^{-2}$ & -1.509(16) \\ \hline
16 & 4&1.320 &40000 & 0.0793(20) & 0.8322(314)~$10^{-2}$ & -1.170(51) \\ \hline
 ~ & ~&1.330 &40000 & 0.0905(19) & 1.0258(339)~$10^{-2}$ & -1.351(31) \\ \hline
 ~ & ~&1.350 &40000 & 0.1111(25) & 1.4542(505)~$10^{-2}$ & -1.556(28) \\ \hline
16 & 8&1.880 &70000 & 0.0441( 3) & 0.2477( 25)~$10^{-2}$ & -1.277(10) \\ \hline
 ~ & ~&1.885 &87500 & 0.0457( 2) & 0.2622( 19)~$10^{-2}$ & -1.329( 9) \\ \hline
 ~ & ~&1.890 &70000 & 0.0463( 3) & 0.2677( 27)~$10^{-2}$ & -1.344(10) \\ \hline
 ~ & ~&1.895 &88000 & 0.0472( 3) & 0.2781( 32)~$10^{-2}$ & -1.348(13) \\ \hline
 ~ & ~&1.900 &70000 & 0.0481( 4) & 0.2869( 35)~$10^{-2}$ & -1.367(13) \\ \hline
 ~ & ~&1.905 &70000 & 0.0499( 4) & 0.3057( 33)~$10^{-2}$ & -1.411(14) \\ \hline
 ~ & ~&1.910 &70000 & 0.0504( 7) & 0.3118( 60)~$10^{-2}$ & -1.411(19) \\ \hline
 ~ & ~&1.915 &70000 & 0.0524( 6) & 0.3329( 58)~$10^{-2}$ & -1.453(18) \\ \hline
 ~ & ~&1.920 &20000 & 0.0528( 7) & 0.3396( 71)~$10^{-2}$ & -1.442(22) \\ \hline
24 & 8&1.880 &80000 & 0.0340( 6) & 0.1483( 44)~$10^{-2}$ & -1.257(27) \\ \hline
 ~ & ~&1.885 &80000 & 0.0353( 6) & 0.1585( 37)~$10^{-2}$ & -1.289(32) \\ \hline
 ~ & ~&1.890 &60000 & 0.0369( 5) & 0.1708( 35)~$10^{-2}$ & -1.350(21) \\ \hline
 ~ & ~&1.895 &94000 & 0.0394( 6) & 0.1892( 46)~$10^{-2}$ & -1.440(24) \\ \hline
 ~ & ~&1.900 &60000 & 0.0404( 4) & 0.1984( 33)~$10^{-2}$ & -1.447(14) \\ \hline
 ~ & ~&1.905 &59000 & 0.0429( 6) & 0.2192( 49)~$10^{-2}$ & -1.514(21) \\ \hline
\end{tabular}
\end{center}
\caption{Summary of parameters and results of the runs at finite temperature.}
\label{tab:FT}
\medskip\noindent
\end{table}

\begin{table}[htb]
\begin{center}
\begin{tabular}{|c|c|c|}   \hline
 $N_\tau$ & $\beta_c$ & $\Delta\beta$ \\ \hline \hline
 2 & 0.100(10) & ~~~~~~~~~ \\ \hline
 4 & 1.332(~4) & 1.232(14) \\ \hline
 8 & 1.886(~6) & 0.554(10) \\ \hline
\end{tabular}
\end{center}
\noindent
\caption{Critical couplings from cumulant crossings.}
\label{tab:betac}
\medskip\noindent
\end{table}

Therefore, we conclude that the finite temperature phase transition
in the PPM with $SU(2)$ gauge symmetry belongs to the same
universality class as the phase transition in the $3d$ Ising model,
and the finite temperature phase transition in the standard Wilson theory.

\section{Heavy quark potential}\label{sec:potential}

We are interested in comparing some physical observables obtained in the
PPM to those obtained with the SWA. Such a comparison will tell us whether
the continuum limit of the two models is the same, as suggested by
universality. To allow a direct comparison we chose equal lattice spacings
(in physical units) by performing (some of) the simulations at the critical
couplings of the deconfinement transition on lattices with the same
temporal extent $N_\tau$ for both actions.

For such a comparison we chose the heavy quark potential, from which one
can extract the string tension. With the recently developed signal
improving techniques \cite{APE_smearing} the heavy quark potential can be
measured quite accurately without the need for very large statistical
samples. The signal improvement consists in replacing the links that make
up the space--like segments of (time--like) Wilson loops with recursively
constructed ``smeared'' links. Besides planar Wilson loops, with on--axis
space--like segments, we also considered Wilson loops with off--axis
space--like segments along the paths (1,1,0) and (2,1,0) and those related
by the cubic symmetry.

Smearing of the space--like links gives a better overlap with the lowest
state in the exponential behavior
\beq
W(\vec R,T) = \sum_i c_i({\vec R}) \exp\{ -V_i (\vec R) \cdot T \}
\eeq
with $V(\vec R) = V_0 (\vec R)$ the heavy quark potential we would like to
find. With a better overlap the effective potential, extracted from
\beq
V_T(\vec R) = \log \left( \frac{W(\vec R,T)}{W(\vec R,T+1)} \right)
 ~ \conlim ~ V(\vec R)
\label{eq:eff_pot}
\eeq
will have an earlier plateau in $T$, where the statistical errors are still
small.

We fit the potential, taken as the effective potential at a given $T$, to
the usual form \cite{fit_form}
\beq
V(\vec R) = V_0 + \sigma R - \frac{e}{R} -
             f \left( G_L(\vec R) - \frac{1}{R} \right)~~~.
\label{eq:pot_fit}
\eeq
Here $G_L$ denotes the lattice Coulomb potential, which takes into account
the lattice artifacts present at smaller distances; it helps in getting good
fits that also include rather small distances. We used fully correlated
fits with the covariance matrix estimated by a bootstrap method. In all
cases, the best fit values obtained in this way did not differ
significantly from those of naive, uncorrelated fits. To select one from
the many possible fits, obtained when varying the range over which the fit
is performed, we define a ``quality'' of the fit as the product of
confidence level times the number of degrees of freedom divided by the
relative error of the string tension. We selected the fits with the highest
quality and list the results in Table \ref{tab:Pot_fits}. The $T$ in the
third column gives the time separation from which the potential was
determined as in eq.~(\ref{eq:eff_pot}).

For the smallest $\beta$ value considered, corresponding to the critical
coupling of a finite temperature system with $N_\tau = 2$ (see the previous
section), rotational invariance of the potential is quite strongly violated.
Thus we made fits over the potential along a principle axis, and along the
(1,1,0) directions separately and list the result in two 
rows
of the Table
\ref{tab:Pot_fits}.  There was no sign of a Coulomb term --- the string
tension term completely dominates the potential --- so we left it out from
the fit.

For the next coupling, $\beta = 1.332$, the critical coupling of the
$N_\tau =4$ finite temperature transition, rotational invariance is already
quite well restored. But, as an additional check we also fit over the
on--axis potential and the potential along direction (1,1,0) separately and
list them in 
rows 
2 and 3 for each $T$.

\begin{table}[htb]
\begin{center}
\begin{tabular}{|l|r|c||l|l|l|l|l|l|l|}    \hline
 $\beta$ & $L$ & T & $V_0$ & $\sigma$ & e & f & range & $\chi^2/dof$ & CL
 \\ \hline \hline
 0.1   &  8 & 1 & 0.125(11) & 0.674( 5) & & & 2.00 - 4.00 & 3.704/1 & 0.025
 \\ \cline{3-10}
       &    & 1 & 0.211( 5) & 0.695( 3) & & & 1.41 - 5.66 & 1.567/2 & 0.536
 \\ \cline{3-10}
       &    & 2 & 0.068( 4) & 0.692( 3) & & & 1.00 - 3.00 & 3.497/1 & 0.030
 \\ \cline{3-10}
       &    & 2 & 0.224(23) & 0.681(16) & & & 1.41 - 5.66 & 0.855/2 & 0.789
 \\ \hline
 1.332 & 12 & 3 & 0.601(10) & 0.1319(15) & 0.275(15) & 0.72(7) & 1.41 - 8.49 &
 7.721/10 & 0.751 \\ \cline{3-10}
       &    & 3 & 0.599(11) & 0.1321(20) & 0.286(13) & & 2.00 - 6.00 &
 0.872/2 & 0.783 \\ \cline{3-10}
       &    & 3 & 0.588( 7) & 0.1343(13) & 0.254( 7) & & 1.41 - 8.49 &
 1.599/3 & 0.784 \\ \cline{3-10}
       &    & 4 & 0.597( 6) & 0.1317(12) & 0.267( 6) & 0.468(14) & 1.41 -
 8.49 & 8.758/10 & 0.619 \\ \cline{3-10}
       &    & 4 & 0.591(19) & 0.1332(36) & 0.276(22) & & 2.00 - 6.00 &
 0.243/2 & 0.975 \\ \cline{3-10}
       &    & 4 & 0.595(11) & 0.1323(23) & 0.261(11) & & 1.41 - 8.49 &
 0.544/3 & 0.982 \\ \cline{3-10}
       &    & 5 & 0.596(10) & 0.1318(23) & 0.266(10) & 0.471(20) & 1.41 -
 8.49 & 5.715/10 & 0.934 \\ \cline{3-10}
       &    & 5 & 0.630(36) & 0.1253(71) & 0.321(42) & & 2.00 - 6.00 &
 0.031/2 & 0.999 \\ \cline{3-10}
       &    & 5 & 0.587(21) & 0.1342(46) & 0.253(21) & & 1.41 - 8.49 &
 1.028/3 & 0.915 \\ \cline{2-10}
       & 16 & 3 & 0.538(21) & 0.1386(21) & 0.109(55) & 1.57(27) & 3.00 -
 11.31 & 15.068/11 & 0.115 \\ \cline{3-10}
       &    & 3 & 0.600( 6) & 0.1326(10) & 0.287( 7) & & 2.00 - 8.00 &
 3.089/4 & 0.627 \\ \cline{3-10}
       &    & 3 & 0.581( 3) & 0.1363( 7) & 0.248( 4) & & 1.41 - 11.31 &
 3.200/5 & 0.781 \\ \cline{3-10}
       &    & 4 & 0.587( 8) & 0.1353(14) & 0.259(12) & 0.694(60) & 2.24 -
 11.31 & 7.626/13 & 0.953 \\ \cline{3-10}
       &    & 4 & 0.590(10) & 0.1347(19) & 0.277(12) & & 2.00 - 8.00 &
 2.039/4 & 0.850 \\ \cline{3-10}
       &    & 4 & 0.577( 6) & 0.1372(13) & 0.244( 6) & & 1.41 - 11.31 &
 1.019/5 & 0.996 \\ \cline{3-10}
       &    & 5 & 0.583( 5) & 0.1361(12) & 0.255( 5) & 0.454(11) & 1.41 -
 11.31 & 11.604/15 & 0.807 \\ \cline{3-10}
       &    & 5 & 0.599(19) & 0.1333(36) & 0.289(22) & & 2.00 - 8.00 &
 1.585/4 & 0.923 \\ \cline{3-10}
       &    & 5 & 0.581(12) & 0.1367(27) & 0.249(12) & & 1.41 - 11.31 &
 2.855/5 & 0.839 \\ \hline
 1.6   & 12 & 3 & 0.569( 2) & 0.0751( 3) & 0.238( 2) & 0.319( 8) & 1.41 -
 8.49 & 9.285/10 & 0.550 \\ \cline{3-10}
       &    & 4 & 0.571( 2) & 0.0747( 4) & 0.240( 3) & 0.322( 9) & 1.41 -
 8.49 & 5.681/10 & 0.936 \\ \cline{3-10}
       &    & 5 & 0.571( 3) & 0.0746( 6) & 0.240( 3) & 0.318(11) & 1.41 -
 8.49 & 4.944/10 & 0.970 \\ \cline{2-10}
       & 16 & 3 & 0.571( 1) & 0.0742( 2) & 0.240( 1) & 0.316( 6) & 1.41 -
 11.31 & 8.662/15 & 0.968 \\ \cline{3-10}
       &    & 4 & 0.568( 2) & 0.0748( 3) & 0.237( 2) & 0.313( 7) & 1.41 -
 11.31 & 6.076/15 & 0.998 \\ \cline{3-10}
       &    & 5 & 0.569( 2) & 0.0746( 5) & 0.237( 3) & 0.307( 9) & 1.41 -
 11.31 & 11.096/15 & 0.847 \\ \cline{3-10}
       &    & 6 & 0.568( 3) & 0.0750( 7) & 0.238( 3) & 0.315(10) & 1.41 -
 11.31 & 4.847/15 & 0.999 \\ \hline
 1.8   & 12 & 3 & 0.555( 2) & 0.0435( 3) & 0.232( 2) & 0.267( 9) & 1.41 -
 8.49 & 3.105/10 & 0.998 \\ \cline{3-10}
       &    & 4 & 0.559( 2) & 0.0423( 4) & 0.236( 3) & 0.273(10) & 1.41 -
 8.49 & 8.076/10 & 0.707 \\ \cline{3-10}
       &    & 5 & 0.562( 3) & 0.0417( 5) & 0.238( 3) & 0.277(10) & 1.41 -
 8.49 & 7.082/10 & 0.822 \\ \hline
\end{tabular}
\end{center}
\caption{Fits to the potential approximants $V_T(R)$.}
\label{tab:Pot_fits}
\medskip\noindent
\end{table}

\begin{table}[htb]
\leftline{Table \ref{tab:Pot_fits} continued}
\begin{center}
\begin{tabular}{|l|r|c||l|l|l|l|l|l|l|}    \hline
 $\beta$ & $L$ & T & $V_0$ & $\sigma$ & e & f & range & $\chi^2/dof$ & CL
 \\ \hline \hline
 1.886 & 12 & 3 & 0.547( 2) & 0.0334( 3) & 0.229( 2) & 0.250( 9) & 1.41 -
 8.49 & 5.761/10 & 0.932 \\ \cline{3-10}
       &    & 4 & 0.564( 4) & 0.0307( 6) & 0.255( 7) & 0.212(17) & 2.00 -
 8.49 & 1.585/9 & 0.999 \\ \cline{3-10}
       &    & 5 & 0.574( 5) & 0.0290( 7) & 0.268( 8) & 0.203(19) & 2.00 -
 8.49 & 5.110/9 & 0.924 \\ \cline{2-10}
       & 16 & 3 & 0.540( 1) & 0.0395( 1) & 0.223( 1) & 0.246( 4) & 1.41 -
 11.31 & 6.390/15 & 0.997 \\ \cline{3-10}
       &    & 4 & 0.547( 2) & 0.0347( 2) & 0.234( 3) & 0.228( 9) & 2.00 -
 11.31 & 4.601/14 & 0.999 \\ \cline{3-10}
       &    & 5 & 0.552( 2) & 0.0339( 3) & 0.241( 4) & 0.225(10) & 2.00 -
 11.31 & 1.661/14 & 1.000 \\ \cline{3-10}
       &    & 6 & 0.546( 2) & 0.0348( 2) & 0.227( 1) & 0.252( 6) & 1.41 -
 11.31 & 10.413/15 & 0.893 \\ \cline{3-10}
       &    & 7 & 0.554( 3) & 0.0336( 4) & 0.244( 5) & 0.217(13) & 2.00 -
 11.31 & 6.270/14 & 0.995 \\ \hline
 2.0   & 12 & 3 & 0.544( 3) & 0.0215( 5) & 0.240( 6) & 0.202(14) & 2.00 -
 8.49 & 1.947/9 & 0.999 \\ \cline{3-10}
       &    & 4 & 0.555( 4) & 0.0193( 5) & 0.255( 6) & 0.194(16) & 2.00 -
 8.49 & 4.759/9 & 0.947 \\ \cline{3-10}
       &    & 5 & 0.565( 8) & 0.0171(10) & 0.267(14) & 0.187(34) & 2.00 -
 8.49 & 6.979/9 & 0.732 \\ \hline
 2.05  & 12 & 3 & 0.543( 3) & 0.0165( 4) & 0.243( 5) & 0.188(13) & 2.00 -
 8.49 & 3.524/9 & 0.990 \\ \cline{3-10}
       &    & 4 & 0.554( 3) & 0.0142( 4) & 0.258( 5) & 0.178(14) & 2.00 -
 8.49 & 4.745/9 & 0.947 \\ \cline{3-10}
       &    & 5 & 0.567( 4) & 0.0117( 5) & 0.275( 6) & 0.238(50) & 2.24 -
 8.49 & 7.526/8 & 0.521 \\ \cline{2-10}
       & 16 & 4 & 0.541( 2) & 0.0188( 2) & 0.245( 3) & 0.235(26) & 2.24 -
 11.31 & 8.927/13 & 0.881 \\ \cline{3-10}
       &    & 5 & 0.543( 2) & 0.0181( 2) & 0.247( 3) & 0.225(29) & 2.24 -
 11.31 & 8.558/13 & 0.906 \\ \cline{3-10}
       &    & 6 & 0.545( 2) & 0.0177( 3) & 0.250( 4) & 0.241(32) & 2.24 -
 9.90 & 9.298/12 & 0.773 \\ \cline{3-10}
       &    & 7 & 0.559( 4) & 0.0161( 4) & 0.278( 8) & 0.231(34) & 2.83 -
 11.31 & 4.970/12 & 0.995 \\ \hline
 2.2   & 12 & 3 & 0.513( 2) & 0.0105( 3) & 0.228( 4) & 0.170(10) & 2.00 -
 8.49 & 4.649/9 & 0.952 \\ \cline{3-10}
       &    & 4 & 0.524( 4) & 0.0083( 5) & 0.242( 6) & 0.219(50) & 2.24 -
 8.49 & 5.601/8 & 0.797 \\ \cline{3-10}
       &    & 5 & 0.543( 9) & 0.0054(10) & 0.274(18) & 0.223(77) & 2.83 -
 8.49 & 3.378/7 & 0.944 \\ \cline{2-10}
       & 16 & 4 & 0.509( 2) & 0.0125( 3) & 0.226( 3) & 0.204(33) & 2.24 -
 11.31 & 7.660/13 & 0.951 \\ \cline{3-10}
       &    & 5 & 0.514( 2) & 0.0115( 3) & 0.232( 5) & 0.212(51) & 2.24 -
 11.31 & 8.938/13 & 0.880 \\ \cline{3-10}
       &    & 6 & 0.525( 5) & 0.0100( 5) & 0.253(11) & 0.217(51) & 2.83 -
 11.31 & 3.751/12 & 0.999 \\ \cline{3-10}
       &    & 7 & 0.531( 6) & 0.0091( 6) & 0.262(12) & 0.234(58) & 2.83 -
 11.31 & 9.678/12 & 0.733 \\ \hline
\end{tabular}
\end{center}
\medskip\noindent
\end{table}

To give an impression of the systematic uncertainty in the determination of
the string tension we list in Table \ref{tab:Pot_fits} the results for
several $T$ (see eq.~(\ref{eq:eff_pot})). For the largest coupling listed,
$\beta = 2.2$, we see that even on the $16^4$ lattice the fit parameters
change rapidly with varying $T$. This indicates that the lattice is too
small.
Indeed, from the MCRG and $T_c$ results we estimate that $\beta=2.2$
is quite close to the critical coupling for the deconfinement transition
with $N_\tau=16$. We therefore regard the results for the coupling $\beta=2.2$
as not trustworthy.
Also for $\beta = 2.05$, even on the $L=16$ lattice, we find no good
convergence of the fit parameters to the potential. Thus, the string
tension extracted for this coupling might still suffer from too small a
lattice size. Keeping this in mind, we will nevertheless show results
for $\beta = 2.05$.

\begin{figure}[htb]
\begin{center}
\vskip -1.0truecm
\leavevmode
\epsfysize=420pt\epsfbox{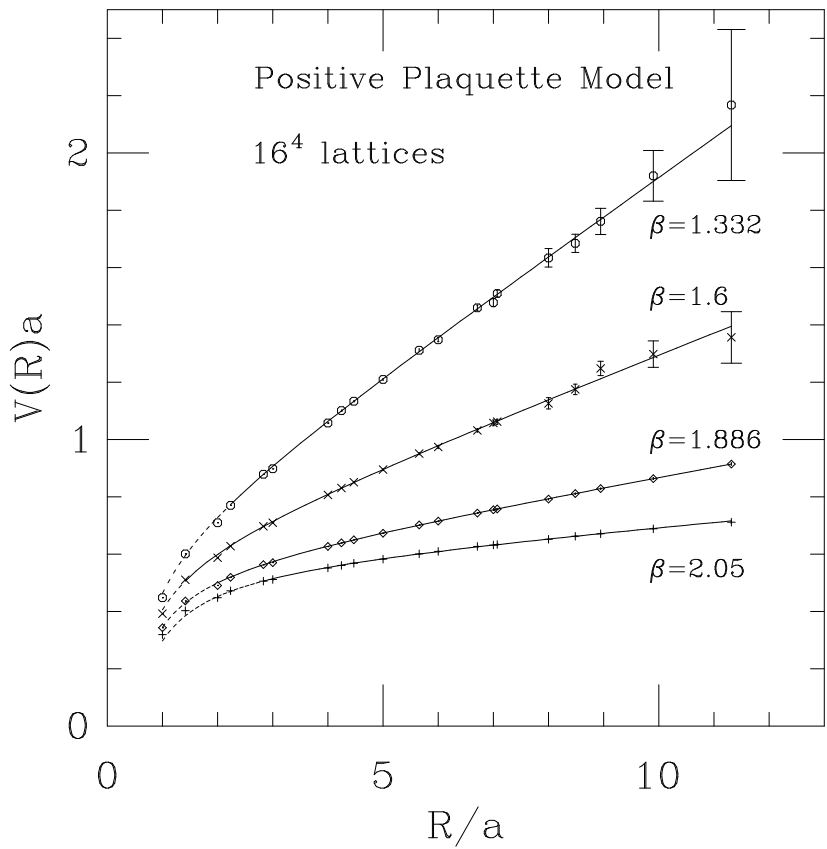}
\end{center}
\vskip -1.0truecm
\caption{The heavy quark potential in the positive plaquette model
         from $16^4$ lattices.}
\label{fig:pot}
\end{figure}

\begin{figure}[htb]
\begin{center}
\vskip -1.0truecm
\leavevmode
\hbox{
\epsfysize=240pt\epsfbox{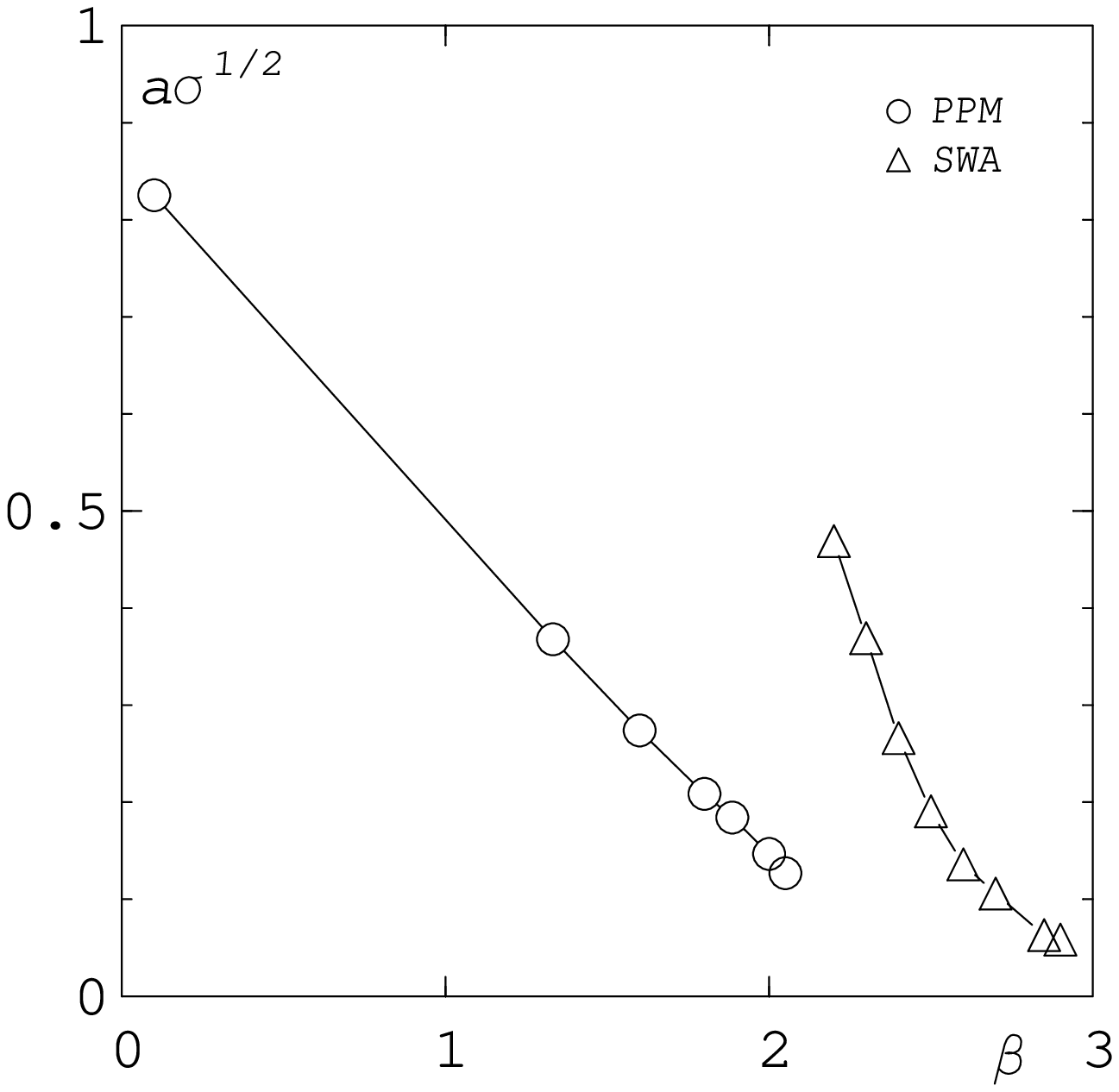}\hskip -1.0truecm
\epsfysize=240pt\epsfbox{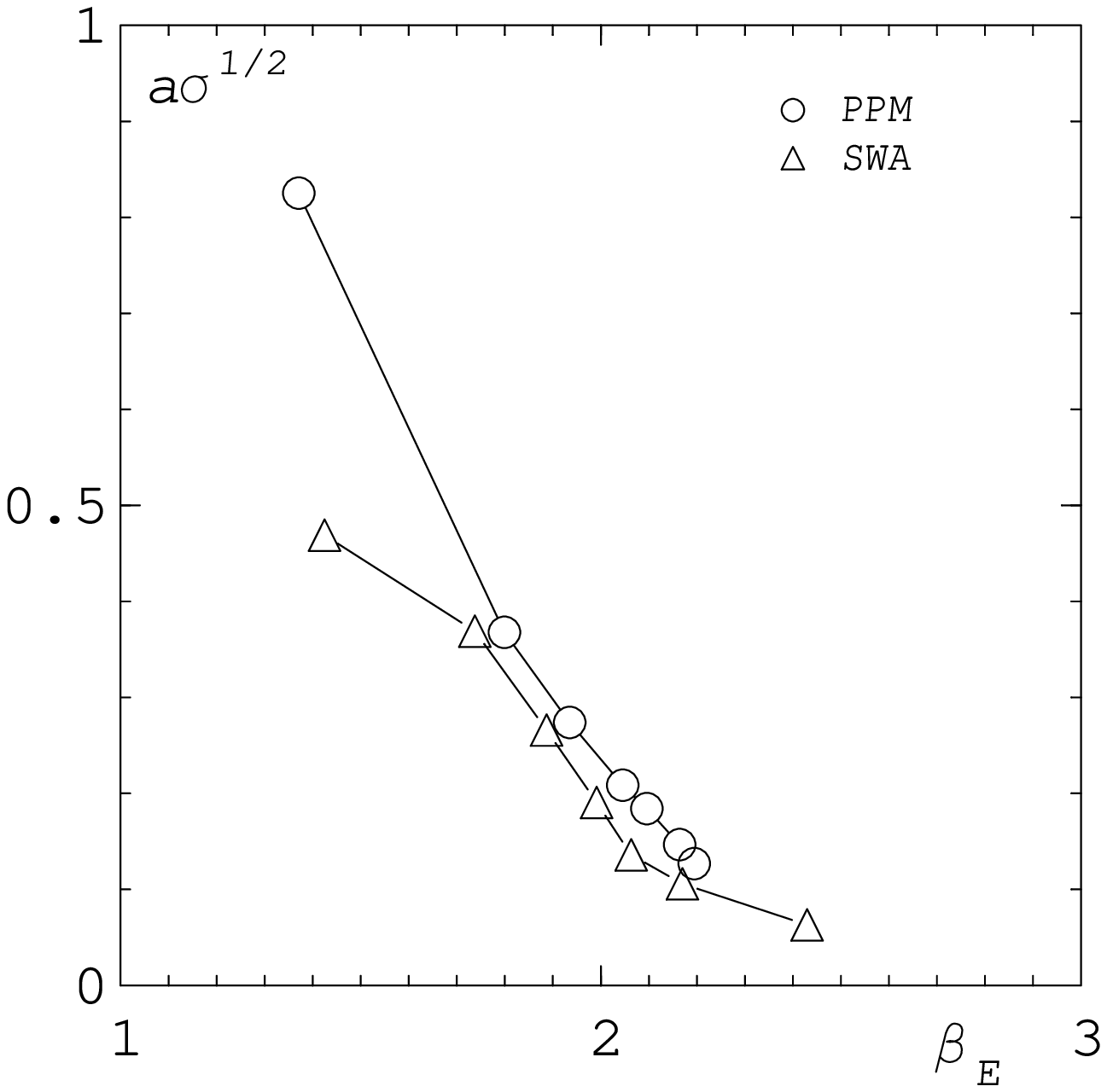}
     }
\end{center}
\vskip -2.0truecm
\caption{A comparison of the string tension in the PPM and for the SWA
         versus the bare lattice coupling (left) and the effective
         coupling $\beta_E$ (right). We compare to SWA results from
         Refs.~\protect\cite{M_T,Mi_P,UKQCD_93}.}
\label{fig:sig}
\end{figure}

For each coupling we chose the fit with the largest confidence level on the
largest lattice in Table \ref{tab:Pot_fits}. The potentials with these
fits, on the $16^4$ lattices, are shown in Fig. \ref{fig:pot}. 
We show $a \sqrt{\sigma}$ versus the bare lattice coupling in
Fig.~\ref{fig:sig}, where for comparison we also included some values for
the SWA. Since the bare couplings for equal lattice spacing in the PPM and
SWA are rather different, the comparison becomes easier, if instead of the
bare coupling, we use some effective coupling which is closer in both
models. We chose the effective coupling 
$\beta_E$~\cite{K_P} for this (see also the Appendix).
The comparison is shown in Fig.~\ref{fig:sig}b, from which we see that now
the behavior of the string tension looks very similar for both the PPM and
SWA once $\beta_E$ becomes larger than a threshold in the region of
$\approx 1.5$.

\section{Glueball measurements}\label{sec:glueball}

Other physical observables in pure glue gauge theories are the glueballs,
in particular their masses. These mass measurements are notoriously
difficult, especially for glueballs other than those with $0^{++}$ and
$2^{++}$ quantum numbers. Since glueballs were not the focal point of our
research, we restricted our measurement to the masses of glueballs with
those two quantum numbers. They can be measured with simple plaquette
correlation functions and thus we did not measure correlation functions
between loops of bigger size and more complicated shape.

As for the measurement of the heavy quark potential some signal improvement
technique is essential for the success of glueball measurements. Both
smearing of links \cite{APE_smearing} and fuzzing \cite{fuzzing} have been
advocated before the ``plaquettes'' for the correlation functions are
constructed. Since we already made smeared (space--like) links for the
potential measurement we decided to try a hybrid of the two methods. We
started with these smeared links and then applied fuzzing to them,
measuring the plaquettes, summed over each time slice for zero--momentum
projection, for each fuzzing level starting with the zeroth, the smeared
links. In the analysis we could then construct the correlation functions at
each fuzzing level as well as cross correlations between plaquettes at
different fuzzing levels.

With $P_{xy}$ being the (fuzzed) plaquette in the $x-y$ plane, etc., we
considered the $0^{++}$ (cubic group $A_1^{++}$) glueball operator
$P_{xy}+P_{xz}+P_{yz}$ and two $2^{++}$ (cubic group $E^{++}$)
glueball operators, $P_{xz}-P_{yz}$ and $2P_{xy}-P_{xz}-P_{yz}$.
We shall refer to these as tensor glueball I and II. In addition we
measured the correlation function between Polyakov lines in the spatial
directions, constructed from the fuzzed links. Their ``mass'' gives another
measurement for the string tension, $m_{Pol} = L \sigma_{Pol}$, with $L$ the
spatial lattice size.
\begin{figure}[htb]
\begin{center}
\vskip -1.0truecm
\leavevmode
\vbox{
\hbox{
\epsfysize=250pt\epsfbox{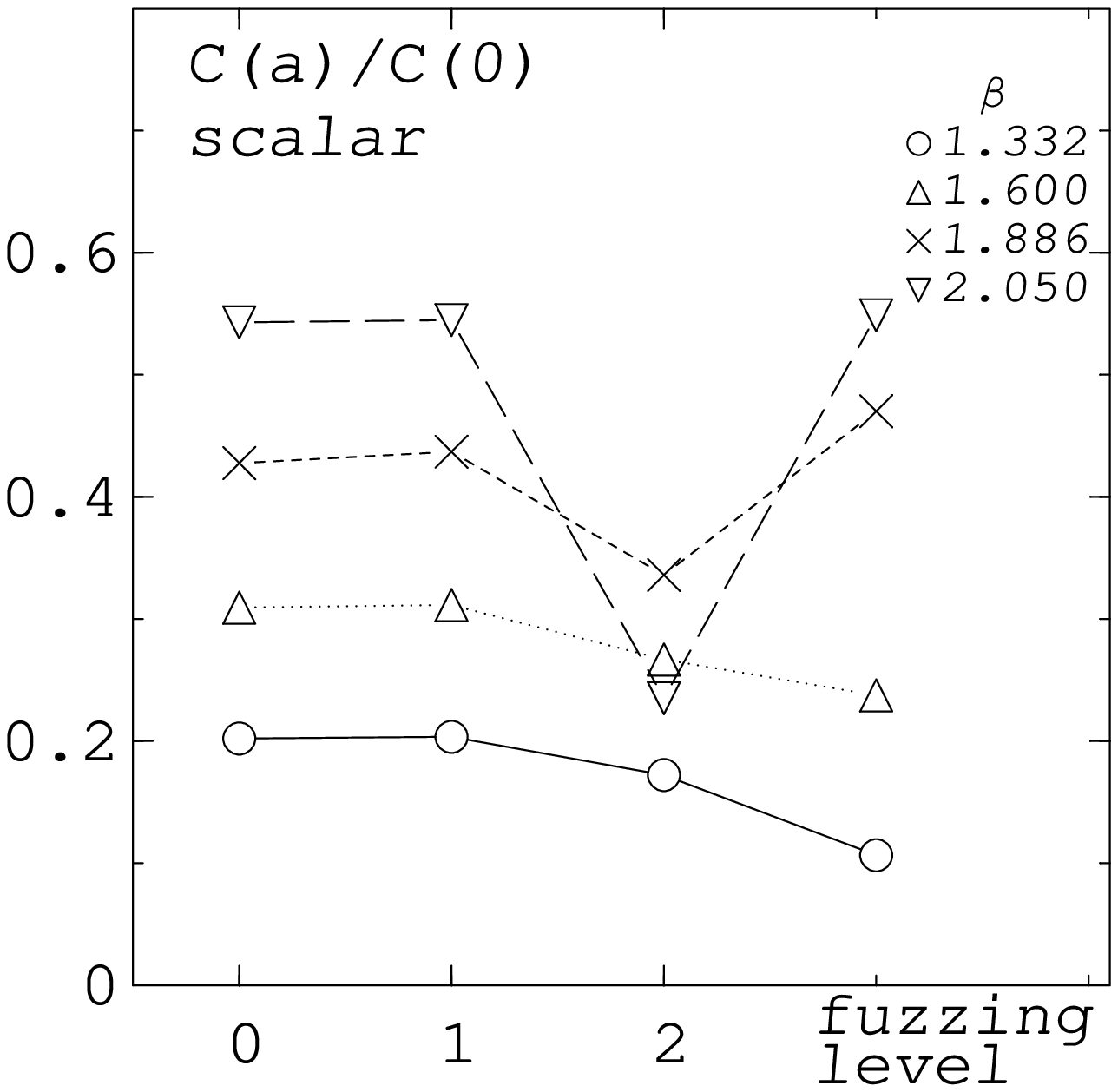} \hskip -1.0truecm
\epsfysize=250pt\epsfbox{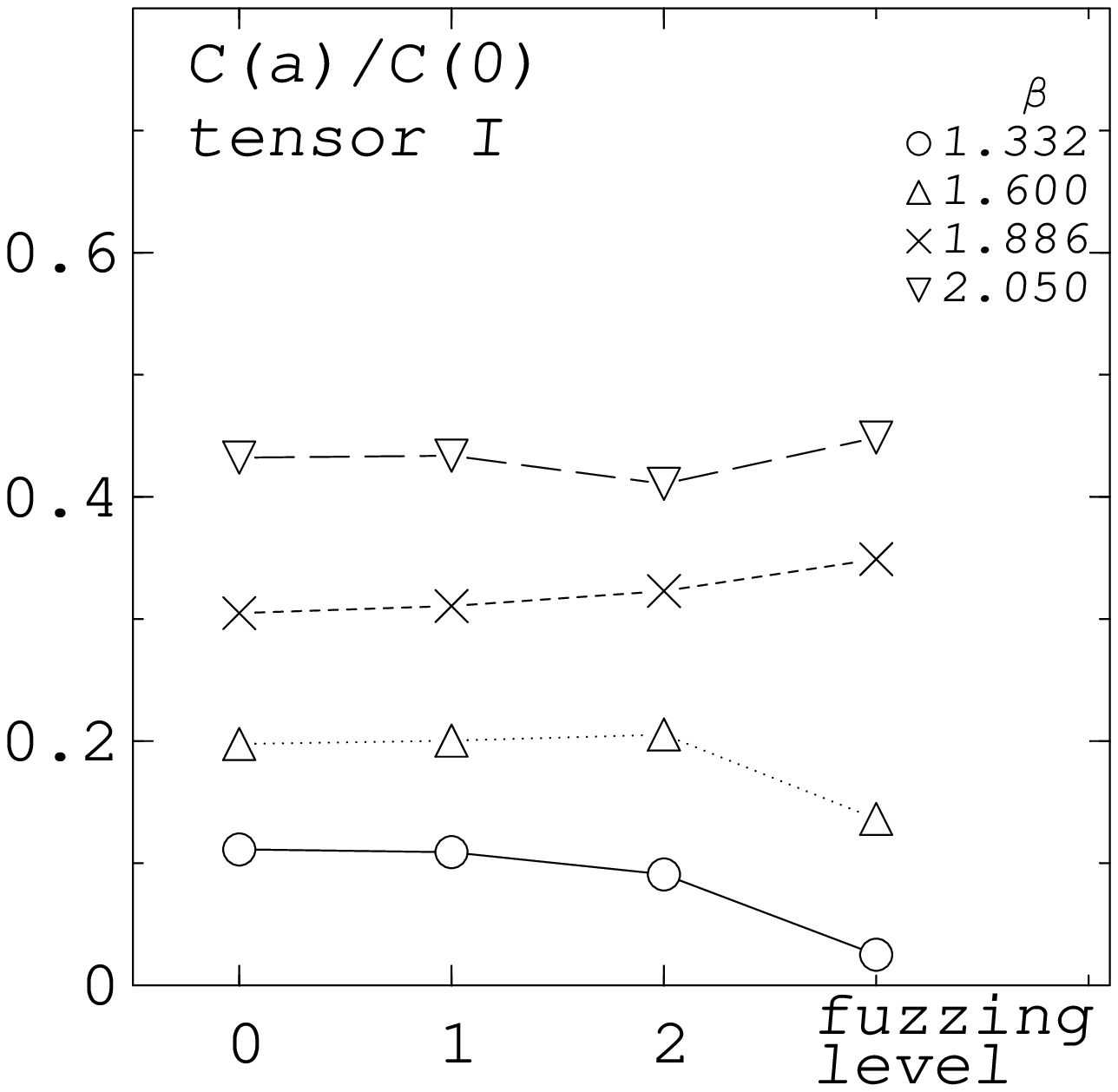}
     } \vskip -3.0truecm
\hbox{
\epsfysize=250pt\epsfbox{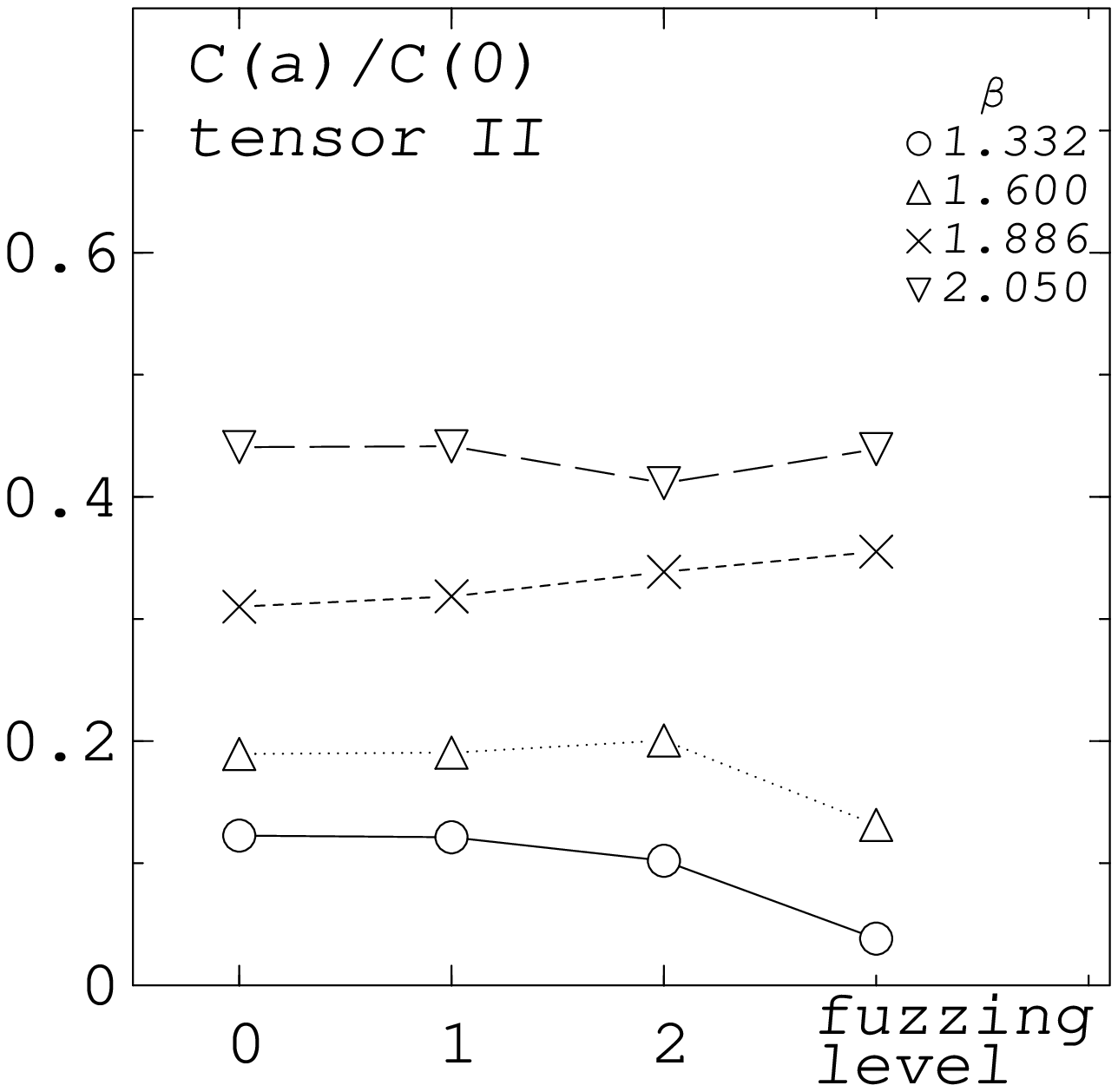}\hskip -1.0truecm
\epsfysize=250pt\epsfbox{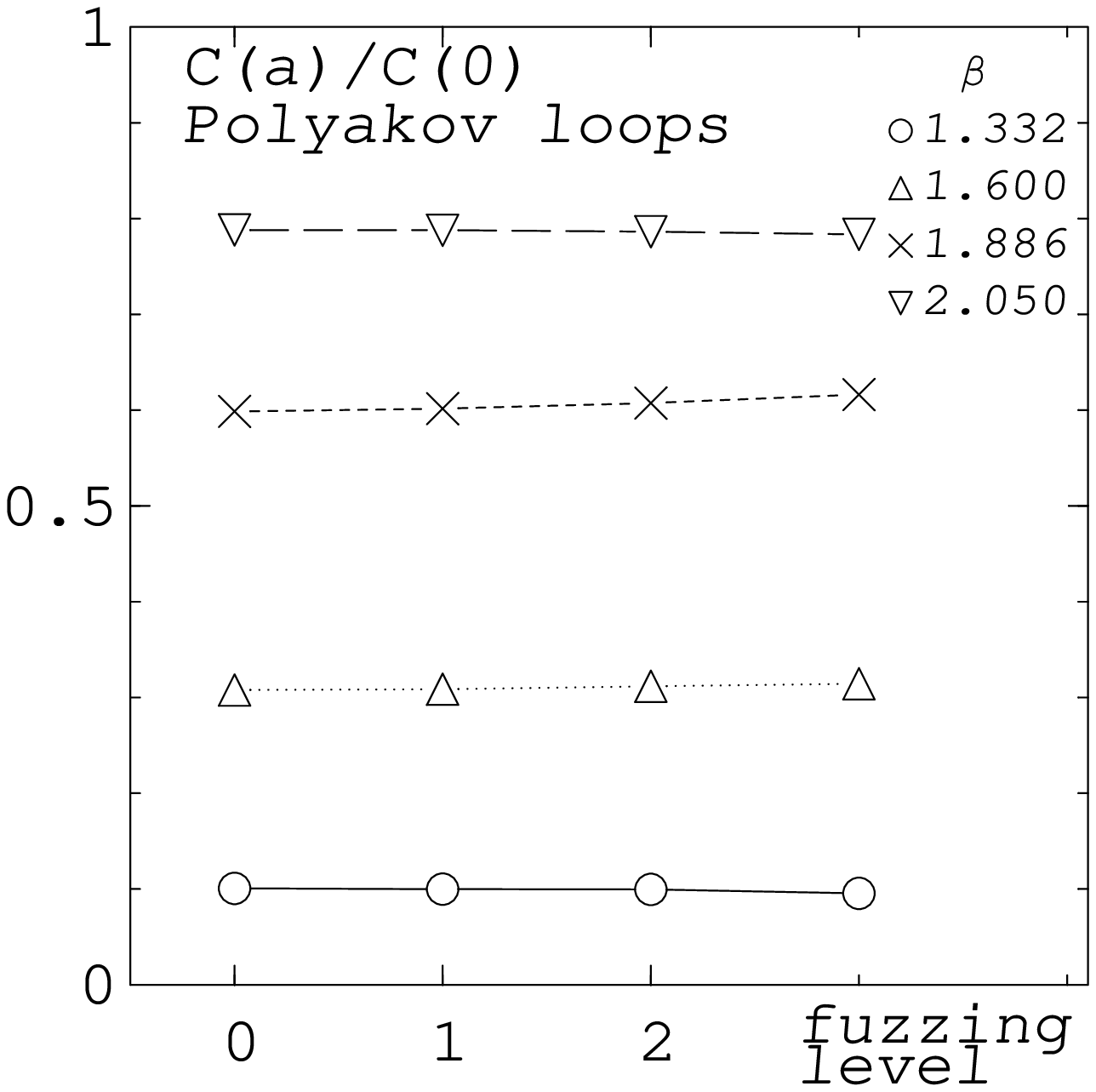}
     }
     }
\end{center}
\vskip -2.2truecm
\caption{Normalized correlation functions for the scalar,
         the tensor I and II glueballs
         and Polyakov loops on $16^4$ lattices.}
\label{fig:v9-12}
\end{figure}
In Fig.~\ref{fig:v9-12} we show the ratios of the correlation functions at
distance $a$ and 0. According to \cite{MiTep} this should give an
indication of the optimal fuzzing level. Except for a somewhat mysterious
dip at fuzzing level 2 for the scalar glueball, this ratio is almost
independent of the fuzzing level. The reason is, that we started, at
fuzzing level 0, with smeared links, which already give a good signal.

Since we measured all cross correlations between the operators with
different fuzzing level, we tried the diagonalization procedure of
\cite{Cross_corr} to improve the quality of the extracted masses.
It did not make any improvement. It appears that the
operators we used are too similar and hence couple almost equally to the
various states in a channel with given quantum numbers. We therefore
extracted the masses simply as effective, or running masses, from the
correlation function at fuzzing level 1. The masses are listed in Table
\ref{tab:glue}.

\begin{table}[htb]
\begin{center}
\begin{tabular}{|l|r|l|||l|l|l|l|}    \hline
 $\beta$ & $L$ & $t$ & $m(0^{++})$ & $m(2^{++})$ I & $m(2^{++})$ II &
 $\sigma_{Pol}(L)$ \\ \hline \hline
 1.332 & 12 & 1 & 1.65( 3) & 2.42( 6) & 2.14( 7) & 0.139( 3) \\ \cline{3-7}
       &    & 2 & 1.26(11) & 2.40(77) & 2.27(51) & 0.156(20) \\ \cline{2-7}
       & 16 & 1 & 1.59( 2) & 2.21( 6) & 2.11( 4) & 0.144( 3) \\ \cline{3-7}
       &    & 2 & 1.38(11) & 2.72(66) & 2.21(41) & 0.166(48) \\ \hline
 1.6   & 12 & 1 & 1.19( 3) & 1.56( 3) & 1.60( 6) & 0.070( 1) \\ \cline{3-7}
       &    & 2 & 0.87( 7) & 1.41(18) & 1.60(19) & 0.073( 4) \\ \cline{2-7}
       & 16 & 1 & 1.17( 3) & 1.61( 4) & 1.66( 5) & 0.073( 1) \\ \cline{3-7}
       &    & 2 & 1.08( 6) & 1.51(17) & 1.58(25) & 0.071( 6) \\ \hline
 1.8   & 12 & 1 & 0.89( 2) & 1.25( 2) & 1.31( 3) & 0.0345(12) \\ \cline{3-7}
       &    & 2 & 0.75( 3) & 1.13( 9) & 1.13(10) & 0.0319(22) \\ \cline{3-7}
       &    & 3 & 0.69( 7) & 1.35(40) & 0.89(17) & 0.0274(19) \\ \hline
 1.886 & 12 & 1 & 0.76( 1) & 1.14( 2) & 1.12( 2) & 0.0244( 9) \\ \cline{3-7}
       &    & 2 & 0.65( 3) & 1.11( 8) & 0.96( 7) & 0.0220(11) \\ \cline{3-7}
       &    & 3 & 0.57( 6) & 0.96(16) & 1.22(19) & 0.0198(12) \\ \cline{2-7}
       & 16 & 1 & 0.83( 1) & 1.17( 3) & 1.14( 3) & 0.0318( 6) \\ \cline{3-7}
       &    & 2 & 0.70( 3) & 1.01( 6) & 1.15( 8) & 0.0321(12) \\ \cline{3-7}
       &    & 3 & 0.69( 6) & 1.15(24) & 2.00(57) & 0.0323(16) \\ \hline
 2.0   & 12 & 1 & 0.73( 2) & 1.04( 2) & 0.99( 2) & 0.0158( 9) \\ \cline{3-7}
       &    & 2 & 0.58( 3) & 0.83( 6) & 0.87( 6) & 0.0136( 9) \\ \cline{3-7}
       &    & 3 & 0.56( 5) & 0.64( 9) & 0.84(12) & 0.0111( 9) \\ \cline{3-7}
       &    & 4 & 0.45( 7) & 0.83(29) & 0.61(24) & 0.0081( 7) \\ \hline
 2.05  & 12 & 1 & 0.69( 2) & 0.92( 2) & 0.93( 2) & 0.0130( 7) \\ \cline{3-7}
       &    & 2 & 0.56( 3) & 0.78( 5) & 0.79( 5) & 0.0106( 8) \\ \cline{3-7}
       &    & 3 & 0.48( 4) & 0.63( 9) & 0.77( 9) & 0.0083( 8) \\ \cline{3-7}
       &    & 4 & 0.35( 5) & 0.41(15) & 0.78(19) & 0.0066( 7) \\ \cline{2-7}
       & 16 & 1 & 0.61( 2) & 0.84( 2) & 0.82( 1) & 0.0149( 5) \\ \cline{3-7}
       &    & 2 & 0.51( 2) & 0.76( 4) & 0.78( 4) & 0.0132( 5) \\ \cline{3-7}
       &    & 3 & 0.44( 3) & 0.72( 9) & 0.68( 8) & 0.0128( 8) \\ \cline{3-7}
       &    & 4 & 0.45( 5) & 0.70(10) & 0.44(10) & 0.0117(10) \\ \hline
\end{tabular}
\end{center}
\caption{Running glueball masses and string tension $\sigma_{Pol}(L)$.}
\label{tab:glue}
\medskip\noindent
\end{table}

While masses have finite size effects that are exponentially suppressed
with the system size, the string tension, as determined from Polyakov line
correlations, has substantial finite size effects. They can be understood in
terms of a picture of a fluctuating string \cite{string}. The fluctuations
create a correction to the string tension, which depends on the lattice
size, and drives $\sigma_{Pol}(L)$ to smaller values
\beq
a^2 \sigma_{Pol}(L) = a^2 \sigma_{Pol}(\infty) -\frac{\pi}{3 N_\sigma^2} .
\label{eq:string_corr}
\eeq
The finite size correction is $0.0073$ for the $12^4$ lattices and $0.0041$
for the $16^4$ lattices.

\begin{figure}[htb]
\begin{center}
\vskip -1.0truecm
\leavevmode
\epsfysize=340pt\epsfbox{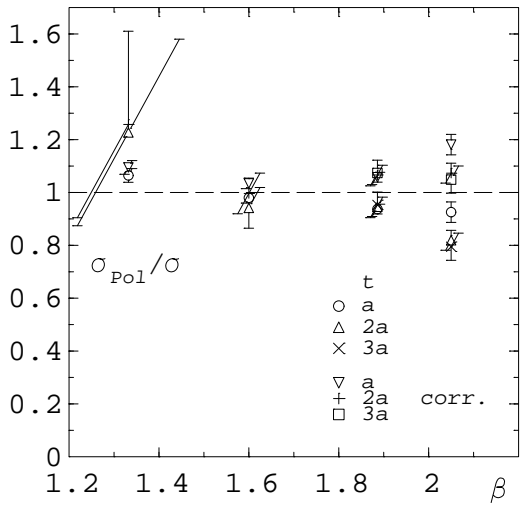}
\end{center}
\vskip -3.0truecm
\caption{Ratio of the string tension from Polyakov lines,
         $\sigma_{Pol}$, and from fits to the heavy
         quark potential, $\sigma$. For $\sigma_{Pol}$ we show both
         the finite volume values and the finite size corrected ones.}
\label{fig:v20}
\end{figure}

We have now two independent determinations of the string tension. We
compare them in Fig.~\ref{fig:v20}, where we show for $\sigma_{Pol}$
both the directly measrued finite volume values, as well as the values
obtained with the finite size corrections of eq.~(\ref{eq:string_corr}).
We see that the finite size corrections become quite substantial for
the larger couplings, and make the agreement better. Of course,
$\sigma_{Pol}$ should be extracted from as large a $t$ as possible,
but generally the last two values agree.

\section{MCRG results}\label{sec:MCRG}

With numerical methods one can not measure the $\beta$--function of a
lattice field theory directly. But, employing Monte Carlo Renormalization
Group (MCRG) methods, we can obtain the ``step $\beta$--function'',
$\Delta \beta(\beta)$, for a given change of scale, in our case by a
factor 2. This step $\beta$--function is the integrated $\beta$--function
over the given finite change of scale.

We employed two different MCRG methods. The first is a real space RG with a
blocking transformation proposed first by Swendsen \cite{Swendsen} that has
a parameter $p$ which can be optimized to reach the renormalized trajectory
after as few blocking steps as possible.  The second is the so called
``ratio method'' in which appropriate ratios of Wilson loops, differing by
a factor 2 in size, are matched. Following \cite{Ratio} we used, besides the
``bare'' ratios, tree--level and one--loop improved ratios in the matching
procedure.

For the real space RG we used ``blocking scheme 1'' of \cite{Blocking},
and followed the optimization procedure described in detail there.
To narrow down the range of the optimization parameter $p$ we made some
trial runs comparing $8^4$ with $4^4$ lattices for each $\beta =
\beta_L$ for which we wanted to find $\Delta \beta$. The optimal value
of $p$ is the one that gives the most consistent matching of all four
observables that we used, the plaquette and the three different shaped
loops of length 6, build from blocked links at the appropriate blocking
level, after the fewest blocking steps. We used linear interpolation in
$1/p$ as in Ref.~\cite{Blocking}. Then we made runs with two values of $p$ that
appeared to bracket the optimal value, comparing $12^4$ with $6^4$ and
$16^4$ with $8^4$ lattices. To be able to do the matching, we made
two runs on the smaller lattices with slightly different couplings,
$\beta_S$, and interpolated linearly between them. For each
of the four observables we obtained in this way a $\Delta\beta =
\beta_L - \beta_S$, with a statistical error determined by jackknife.
We interpolated these to the optimal $p$ values and then averaged them.
The results are listed in Table \ref{tab:Block_res}. The errors quoted
include the statistical errors, as well as systematic errors describing
the spread of $\Delta \beta$ obtained from the different observables.
The columns without the label ``extr.'' give the results of the matching
of blocking level $n$ on the larger lattice with blocking level $n-1$ on
the smaller lattice. The columns with the label ``extr.'' give the
results after extrapolating from blocking levels $n$ and $n-1$ to an
infinite number of blockings, taking into account only the leading
irrelevant eigenoperator of the blocking RG transformation with
eigenvalue $1/4$ \cite{Blocking},
\beq
\Delta \beta^{(\infty)} = \frac{1}{3} \left[ 4 \Delta \beta^{(n)} -
\Delta \beta^{(n-1)} \right] .
\label{eq:block_extr}
\eeq

\begin{table}[htb]
\begin{center}
\begin{tabular}{|l|r||l|l|l|l|}    \hline
 $\beta_l$ & $L$ & 2 & 2, extr. & 3 & 3, extr. \\ \hline \hline
 1.332 &  8 & 1.17(4) & 1.17(4) & & \\ \cline{2-6}
       & 12 & 1.26(6) & 1.23(4) & & \\ \cline{2-6}
       & 16 & 1.26(5) & 1.23(3) & 1.30(5) & 1.29(6) \\ \hline
 1.6   &  8 & 0.763(15) & 0.765(22) & & \\ \cline{2-6}
       & 12 & 0.773(15) & 0.778(18) & & \\ \cline{2-6}
       & 16 & 0.774(15) & 0.779(18) & 0.751(14) & 0.743(16) \\ \hline
 1.8   &  8 & 0.617( 8) & 0.637(10) & & \\ \cline{2-6}
       & 12 & 0.606( 9) & 0.621(14) & & \\ \hline
 1.886 &  8 & 0.660(12) & 0.649(13) & & \\ \cline{2-6}
       & 12 & 0.585( 9) & 0.590(14) & & \\ \cline{2-6}
       & 16 & 0.593(13) & 0.594(14) & 0.585( 6) & 0.571( 4) \\ \hline
 2.0   &  8 & 0.610(10) & 0.606(10) & & \\ \cline{2-6}
       & 12 & 0.570( 9) & 0.556(11) & & \\ \hline
 2.05  &  8 & 0.573( 9) & 0.569( 9) & & \\ \cline{2-6}
       & 12 & 0.519( 5) & 0.530(12) & & \\ \cline{2-6}
       & 16 & 0.518( 6) & 0.528(12) & 0.495( 3) & 0.487( 3) \\ \hline
 2.2   &  8 & 0.526( 8) & 0.527( 7) & & \\ \cline{2-6}
       & 12 & 0.511( 8) & 0.507(10) & & \\ \cline{2-6}
       & 16 & 0.512( 6) & 0.510(11) & 0.490( 4) & 0.482( 5) \\ \hline
 2.4   &  8 & 0.475(12) & 0.485( 4) & & \\ \cline{2-6}
       & 12 & 0.457( 6) & 0.461( 7) & & \\ \hline
 2.6   &  8 & 0.418(10) & 0.427( 3) & & \\ \cline{2-6}
       & 12 & 0.408( 6) & 0.415( 5) & & \\ \cline{2-6}
       & 16 & 0.411( 5) & 0.418( 6) & 0.408( 3) & 0.406( 4) \\ \hline
\end{tabular}
\end{center}
\caption{Blocking MCRG results: the columns with label ``$n$'' contain
         the results from matching after $n$ and $n-1$ blocking steps.
         The columns with label ``$n$, extr.'' contain the results from
         extrapolating $\Delta \beta$ to an infinite number of blocking
         steps according to eq.~(\protect\ref{eq:block_extr}).}
\label{tab:Block_res}
\medskip\noindent
\end{table}

To implement the ratio method we measured planar Wilson loops with the
variance reduced by the method of Parisi, Petronzio and Rapuano,
\cite{PPR}. Since, for the PPM we do not have a closed formula to compute
the improved links, we calculated them with typically 20 hits. To increase
the number of Wilson loops from which we can build ratios, especially for
the smaller lattices, we measured Wilson loops up to distances $L/2+2$.
Since the finite size effects on the large and small lattices to be matched
are the same this is acceptable for the ratio method. We would not advocate
using Wilson loops larger than $L/2$ to extract, say, the potential.

Each Wilson loop ratio considered provides a $\Delta \beta$, again infered
from two $\beta$ values on the smaller lattices by interpolation, with an
error determined by jackknife. The results from ``bare'' ratios are
afflicted by lattice artifacts, especially if the ratio contains small
Wilson loops. These artifacts can be corrected for in perturbation theory
by building appropriate linear combinations that give the correct result in
perturbation theory at tree or one--loop level \cite{Ratio}. Since the
perturbation expansion of the PPM and the theory with SWA are identical ---
the removal of negative plaquettes is a non--perturbative procedure --- the
perturbative Wilson loops of \cite{PertLoop} can be used to construct the
improved ratios.

Typically the improvement shows up in a smaller variance of the results
from many different ratios. The results listed in Table \ref{tab:Ratio_res}
represent the average $\Delta \beta$ over all ratios satisfying certain
``cuts'' \cite{Ratio}. The cuts, given in sizes appropriate for the smaller
of the two lattices matched, are (i) maximal length $R_{max}$ of the sides
of a Wilson loop (since we measured loops up to size $L/2+2$) (ii) the
minimal area $A_{min}$ of Wilson loops included (to cut off small distance
lattice artefacts), (iii) area difference $\Delta A_{min}$ in numerator
and denominator of a ratio (the larger this difference is, the faster the
ratio varies with the coupling and the easier it is to find a good matching)
and (iv) maximal statistical error $\delta_{max}$ in $\Delta\beta$ from a
ratio. The errors quoted in Table \ref{tab:Ratio_res} are the variances
over the ratios that passed all cuts.

\begin{table}[htb]
\begin{center}
\begin{tabular}{|l|r||l|l|l|l|l|l|l|}    \hline
 $\beta_l$ & $L$ & basic & tree--level & 1--loop & $R_{max}$ & $\delta_{max}$ &
 $A_{min}$ & $\Delta A_{min}$ \\ \hline \hline
 1.332 &  8 & 1.37(11) & 1.28( 6) & 1.27( 5) & 3 & 0.200 &
 1 & none \\ \cline{2-9}
       & 12 & 1.37(17) & 1.19(12) & 1.21( 7) & 4 & 0.200 &
 2 & 0 \\ \cline{3-9}
       &    & 1.42(16) & 1.18(16) & 1.33(17) & 3 & 0.200 &
 1 & none \\ \cline{2-9}
       & 16 & 1.34( 9) & 1.24( 4) & 1.22( 3) & 5 & 0.150 &
 2 & 1 \\ \cline{3-9}
       &    & 1.40( 7) & 1.31( 6) & 1.24( 2) & 4 & 0.200 &
 2 & 0 \\ \hline
 1.6   &  8 & 0.83( 9) & 0.72(6) & 0.74(6) & 3 & 0.050 &
 1 & none \\ \cline{2-9}
       & 12 & 0.85( 9) & 0.79(5) & 0.735(16) & 4 & 0.030 &
 2 & 0 \\ \cline{3-9}
       &    & 0.81(10) & 0.74(7) & 0.76(8) & 3 & 0.020 &
 1 & none \\ \cline{2-9}
       & 16 & 0.84( 6) & 0.78(3) & 0.771(26) & 5 & 0.030 &
 2 & 1 \\ \cline{3-9}
       &    & 0.86( 9) & 0.81(6) & 0.775(27) & 4 & 0.050 &
 2 & 0 \\ \hline
 1.8   &  8 & 0.67( 7) & 0.60(3) & 0.63(4) & 3 & 0.030 &
 1 & none \\ \cline{2-9}
       & 12 & 0.70( 6) & 0.62(3) & 0.582( 9) & 4 & 0.020 &
 2 & 0 \\ \cline{3-9}
       &    & 0.62(10) & 0.56(5) & 0.59(6) & 3 & 0.020 &
 1 & none \\ \hline
 1.886 &  8 & 0.63(8) & 0.58(3) & 0.60(4) & 3 & 0.050 &
 1 & none \\ \cline{2-9}
       & 12 & 0.65(5) & 0.59(3) & 0.552( 5) & 4 & 0.020 &
 2 & 0 \\ \cline{3-9}
       &    & 0.60(9) & 0.54(4) & 0.557(42) & 3 & 0.020 &
 1 & none \\ \cline{2-9}
       & 16 & 0.63(5) & 0.57(2) & 0.561(16) & 5 & 0.015 &
 2 & 1 \\ \cline{3-9}
       &    & 0.66(6) & 0.58(3) & 0.557( 7) & 4 & 0.015 &
 2 & 0 \\ \hline
 2.0   &  8 & 0.65( 9) & 0.54(4) & 0.55(6) & 3 & 0.050 &
 1 & none \\ \cline{2-9}
       & 12 & 0.59( 6) & 0.56(4) & 0.518( 7) & 4 & 0.020 &
 2 & 0 \\ \cline{3-9}
       &    & 0.53(10) & 0.49(5) & 0.52(5) & 3 & 0.020 &
 1 & none \\ \hline
 2.05  &  8 & 0.59(12) & 0.49(4) & 0.51(6) & 3 & 0.050 &
 1 & none \\ \cline{2-9}
       & 12 & 0.58( 8) & 0.55(4) & 0.498( 9) & 4 & 0.020 &
 2 & 0 \\ \cline{3-9}
       &    & 0.51(11) & 0.47(4) & 0.49(6) & 3 & 0.020 &
 1 & none \\ \cline{2-9}
       & 16 & 0.55( 6) & 0.516(24) & 0.507(12) & 5 & 0.010 &
 2 & 1 \\ \cline{3-9}
       &    & 0.60( 6) & 0.533(24) & 0.512( 8) & 4 & 0.010 &
 2 & 0 \\ \hline
\end{tabular}
\end{center}
\caption{Ratio method MCRG results: the two lines for sizes 12 and 16
         represent different cuts on the Wilson loops and ratios
         considered. The various cuts are described in the text.}
\label{tab:Ratio_res}
\medskip\noindent
\end{table}

\begin{table}[htb]
\begin{center}
\leftline{Table \ref{tab:Ratio_res} continued}
\vspace{0.5cm}
\begin{tabular}{|l|r||l|l|l|l|l|l|l|}    \hline
 $\beta_l$ & $L$ & basic & tree-level & 1-loop & $R_{max}$ & $\delta_{max}$ &
 $A_{min}$ & $\Delta A_{min}$ \\ \hline \hline
 2.2   &  8 & 0.60( 6) & 0.46(3) & 0.52(5) & 3 & 0.050 &
 1 & none \\ \cline{2-9}
       & 12 & 0.59( 6) & 0.52(5) & 0.460( 5) & 4 & 0.020 &
 2 & 0 \\ \cline{3-9}
       &    & 0.53(11) & 0.43(5) & 0.47(5) & 3 & 0.020 &
 1 & none \\ \cline{2-9}
       & 16 & 0.53( 5) & 0.500(19) & 0.494(10) & 5 & 0.010 &
 2 & 1 \\ \cline{3-9}
       &    & 0.57( 7) & 0.516(25) & 0.493( 5) & 4 & 0.015 &
 2 & 0 \\ \hline
 2.4   &  8 & 0.56(12) & 0.47(3) & 0.48(5) & 3 & 0.050 &
 1 & none \\ \cline{2-9}
       & 12 & 0.57( 7) & 0.49(4) & 0.433( 8) & 4 & 0.020 &
 2 & 0 \\ \cline{3-9}
       &    & 0.49(12) & 0.38(6) & 0.43(5) & 3 & 0.020 &
 1 & none \\ \hline
 2.6   &  8 & 0.48(11) & 0.42(2) & 0.44(4) & 3 & 0.050 &
 1 & none \\ \cline{2-9}
       & 12 & 0.53( 6) & 0.44(4) & 0.394( 6) & 4 & 0.020 &
 2 & 0 \\ \cline{3-9}
       &    & 0.45(12) & 0.34(5) & 0.40(5) & 3 & 0.020 &
 1 & none \\ \cline{2-9}
       & 16 & 0.48( 4) & 0.428(20) & 0.419( 9) & 5 & 0.010 &
 2 & 1 \\ \cline{3-9}
       &    & 0.50( 4) & 0.441(17) & 0.422( 4) & 4 & 0.010 &
 2 & 0 \\ \hline
\end{tabular}
\end{center}
\medskip\noindent
\end{table}
\begin{figure}[htb]
\begin{center}
\vskip -1.0truecm
\leavevmode
\hbox{
\epsfysize=260pt\epsfbox{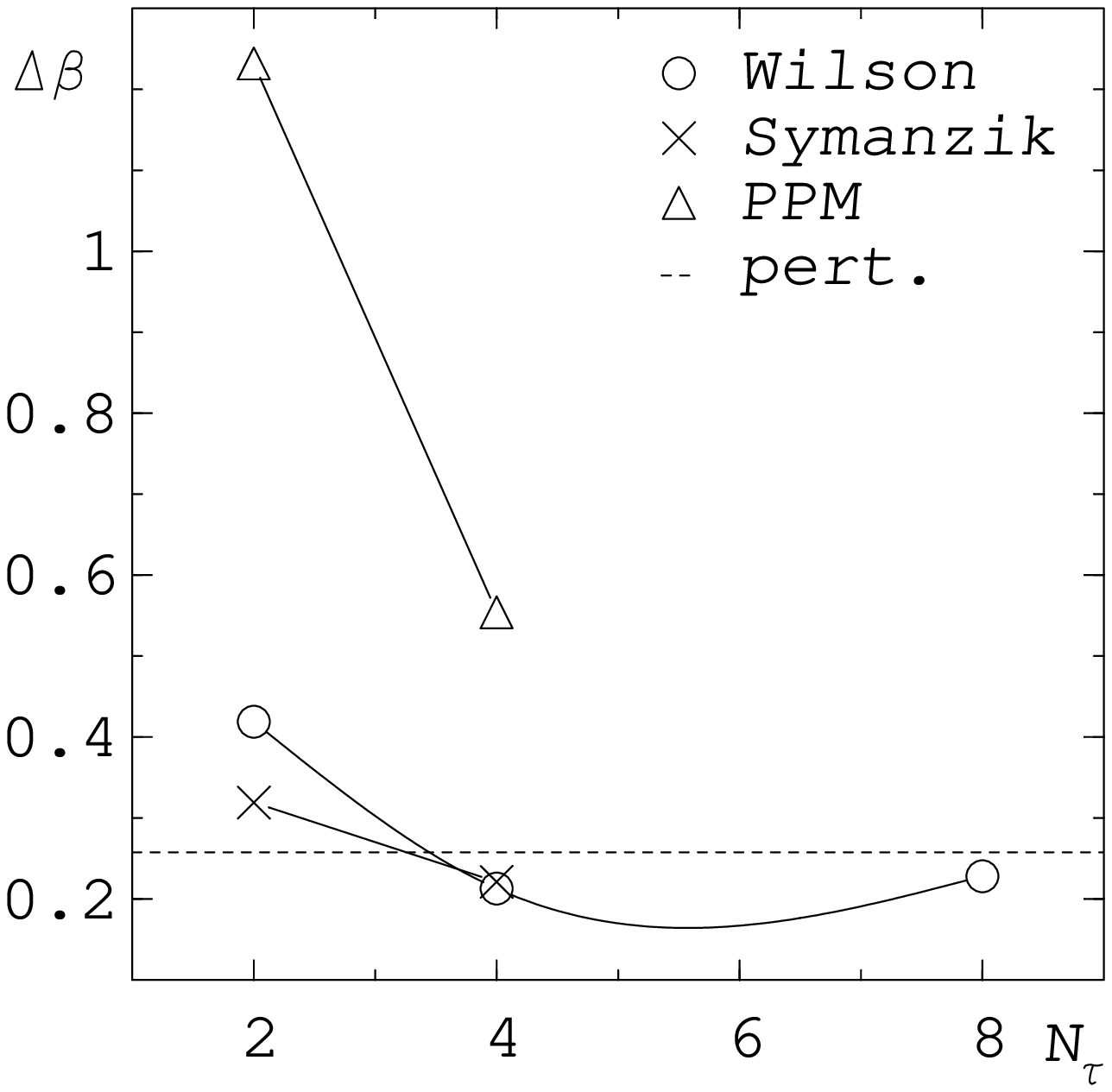} \hskip -0.6truecm
\epsfysize=260pt\epsfbox{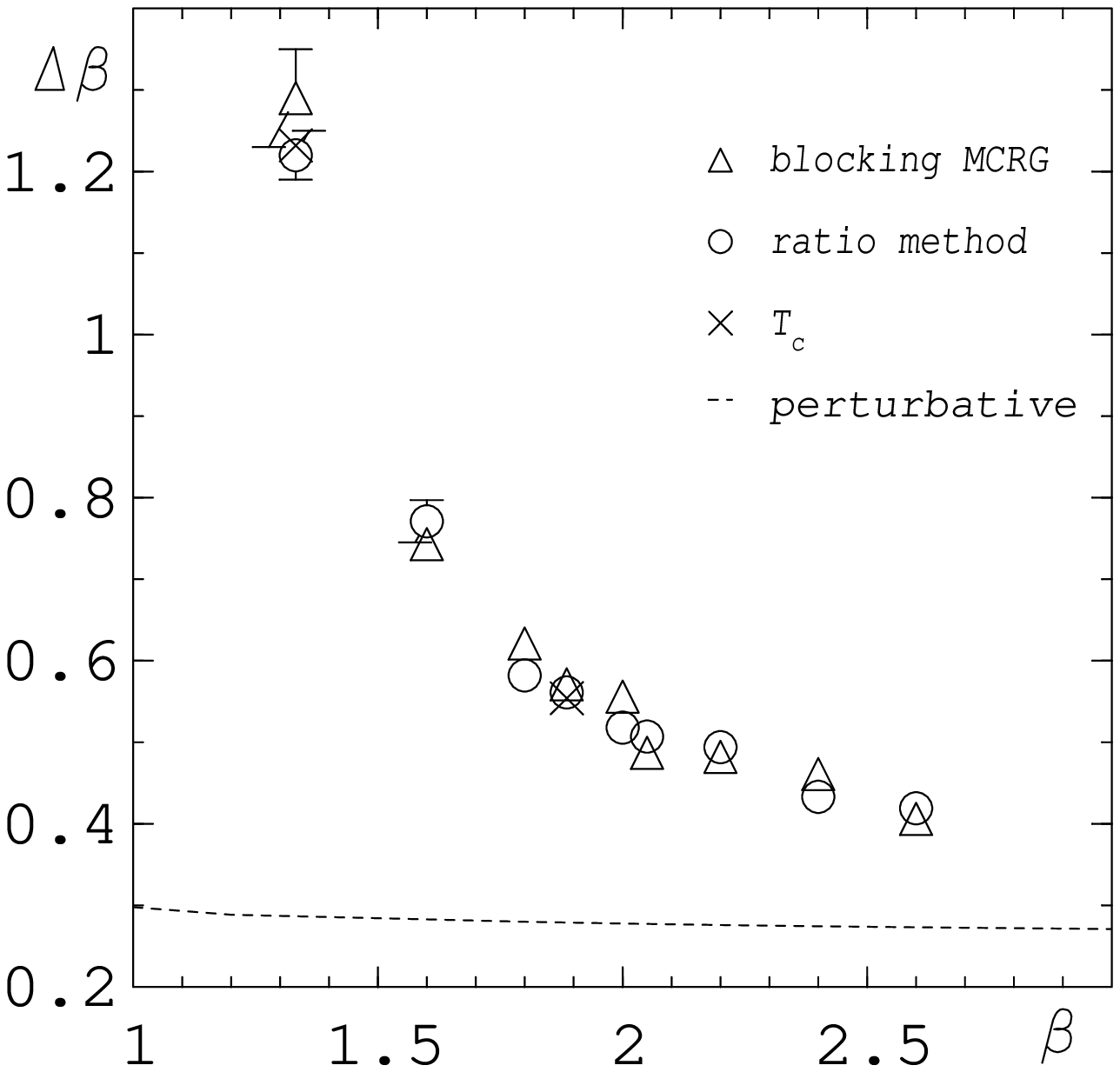}
     }
\end{center}
\vskip -2.0truecm
\caption{The discrete $\beta$--function from measurements of $T_c$ and
         from an MCRG analysis. The figure on the left shows a comparison
         of the SWA and the PPM as well as a Symanzik improved
         action~\protect\cite{Pisa}.
         In the figure on the right we compare the ``step $\beta$--function''
         $\Delta\beta$ obtained by various methods. Data from the largest
         lattice and/or highest blocking level were used.}
\label{fig:dbeta}
\end{figure}
Our results for $\Delta\beta$ from the determination of $T_c$ and from the
MCRG analysis are shown in Fig. \ref{fig:dbeta}. We see that the dip
produced by the SWA is completely removed. The function is monotonous in
the whole range accessible to our simulations. It comes down steeply from
above at small $\beta$ values and converges rather slowly to the
pertubative limit. A comparison at fixed $N_\tau$, \ie, equal lattice
spacing, indicates that the convergence in the PPM is even slower than
with the SWA.
It should be noted that, since the perturbative $\beta$--functions in the
PPM and with SWA are identical, the approach of $\Delta\beta(\beta)$ to
the 2--loop form eventually, for very large $\beta$, has to be the same.
Depending on the sign of the order $1/\beta$ contribution, which
contains the 3--loop coefficient of the $\beta$--function, it comes from
above or below. Thus for either the PPM or the SWA $\Delta\beta(\beta)$
has to cross the 2--loop curve and have another small ``dip'' or ``hill''.

\section{Scaling and asymptotic scaling}\label{sec:scaling}

\begin{figure}[htb]
\begin{center}
\vskip -1.0truecm
\leavevmode
\hbox{
\epsfysize=260pt\epsfbox{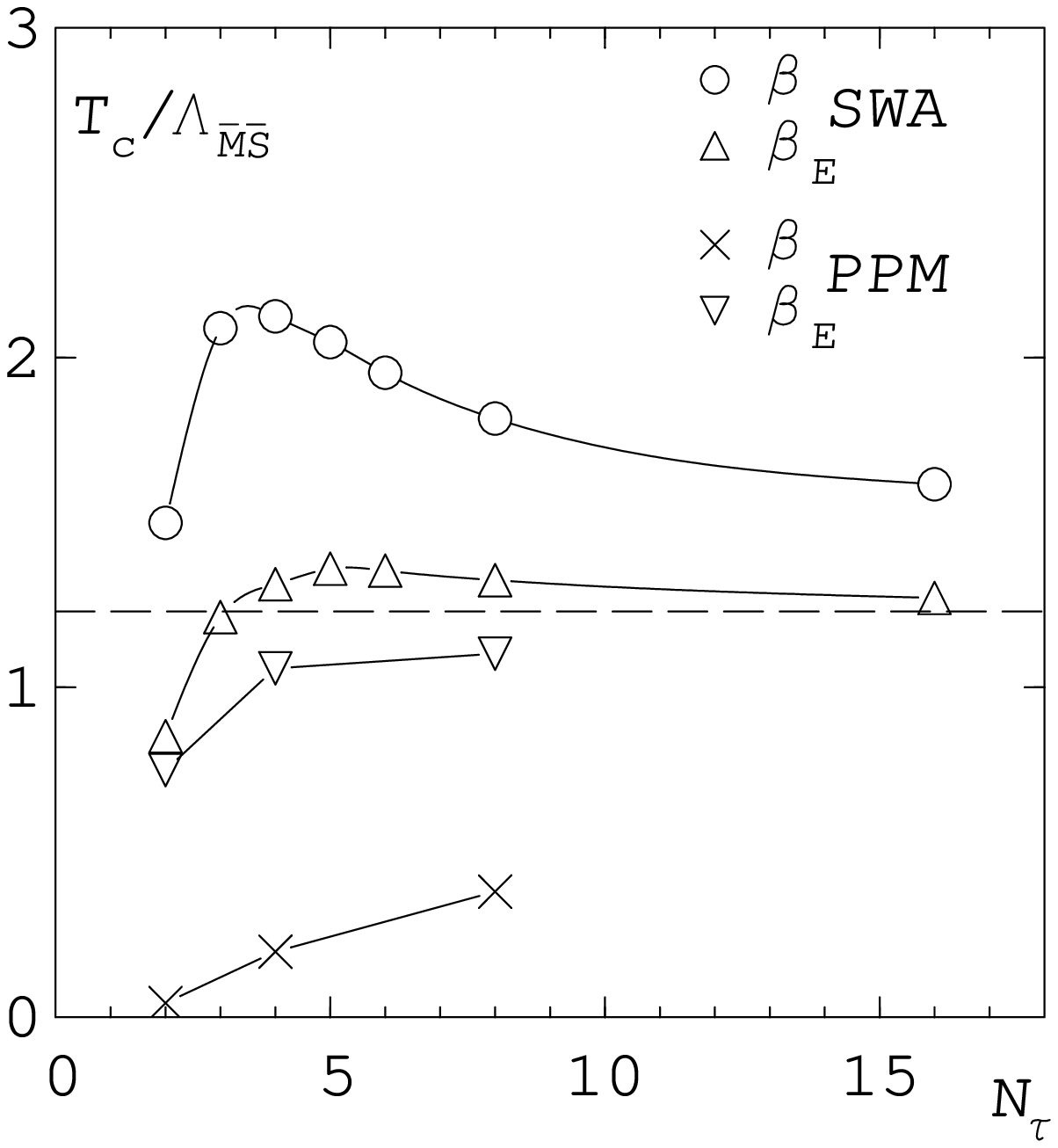}\hskip -1.0truecm
\epsfysize=260pt\epsfbox{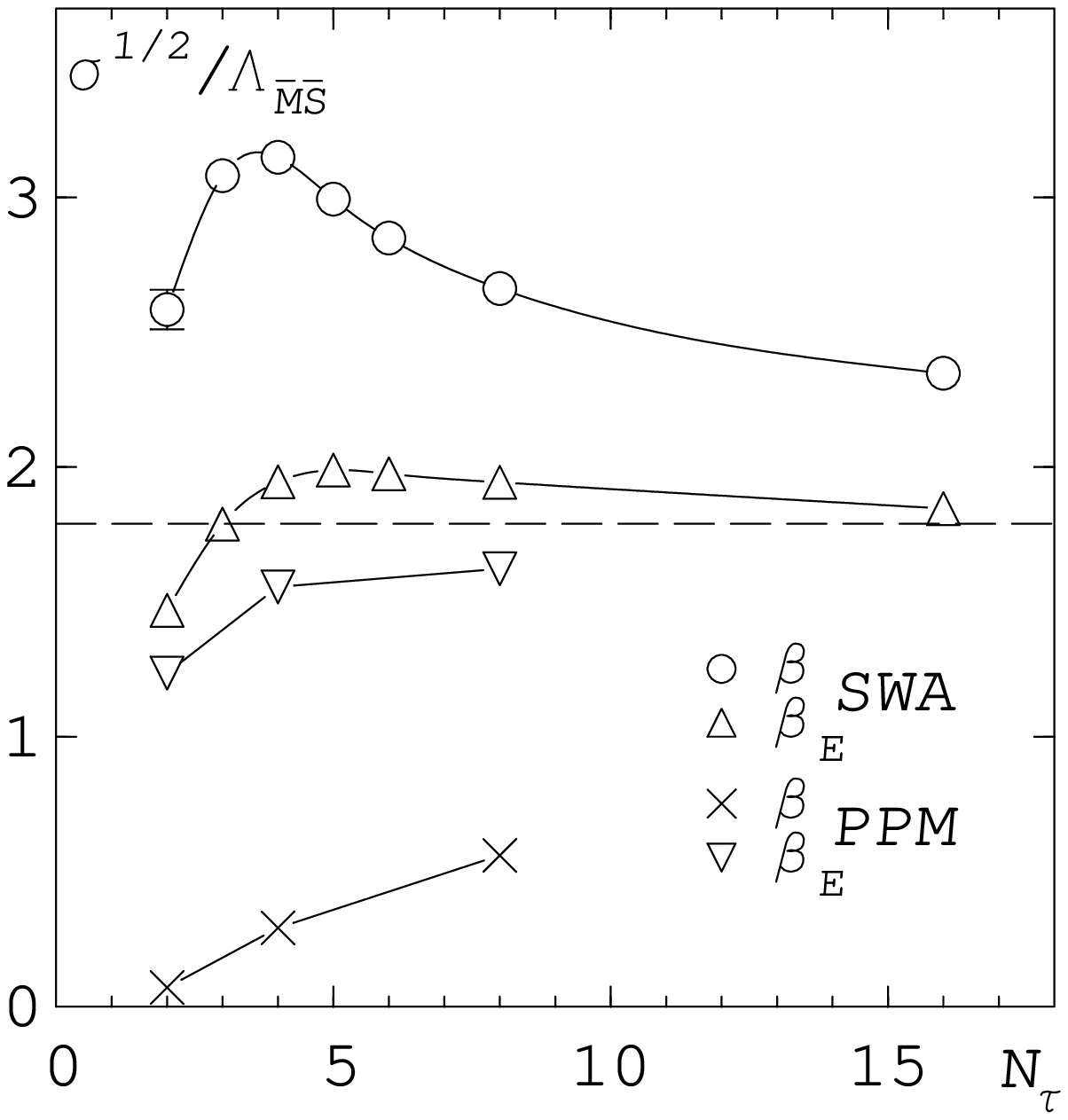}
     }
\end{center}
\vskip -2.0truecm
\caption{Asymptotic scaling of the critical temperature and the
         string tension from fits of the heavy quark potential
         with the bare ($\beta$) and an effective coupling ($\beta_E$).
         The dashed lines indicate the continuum values as determined
         in Ref.~\protect\cite{scaling}.}
\label{fig:j1}
\end{figure}

The results of the last section tell us, that, in the range of couplings
accessible to numerical simulations with moderate resources, we are still
rather far from asymptotic scaling, especially as a function of the bare
lattice coupling. This can be seen in Table \ref{tab:AS} and Fig.
\ref{fig:j1} where we show the critical temperature converted to units of 
$\Lambda_{\overline{\mbox{MS}}}$ using the perturbative 2--loop relation.
Besides giving the values for the bare lattice coupling we also considered an
effective coupling scheme $\beta_E$ \cite{K_P} (see also the Appendix).

\begin{figure}[htb]
\begin{center}
\vskip -1.0truecm
\leavevmode
\hbox{
\epsfysize=280pt\epsfbox{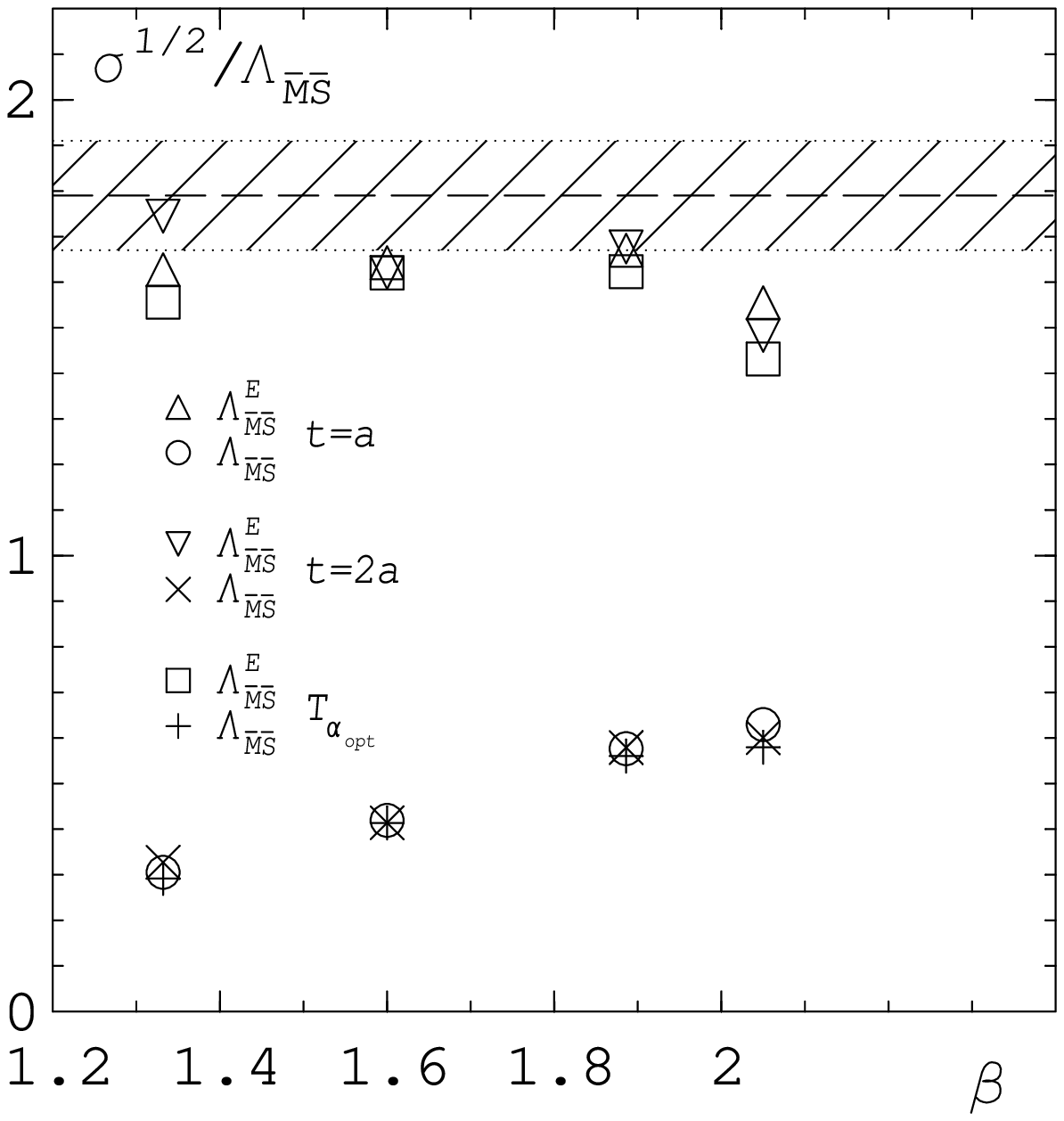} \hskip -0.6truecm
\epsfysize=280pt\epsfbox{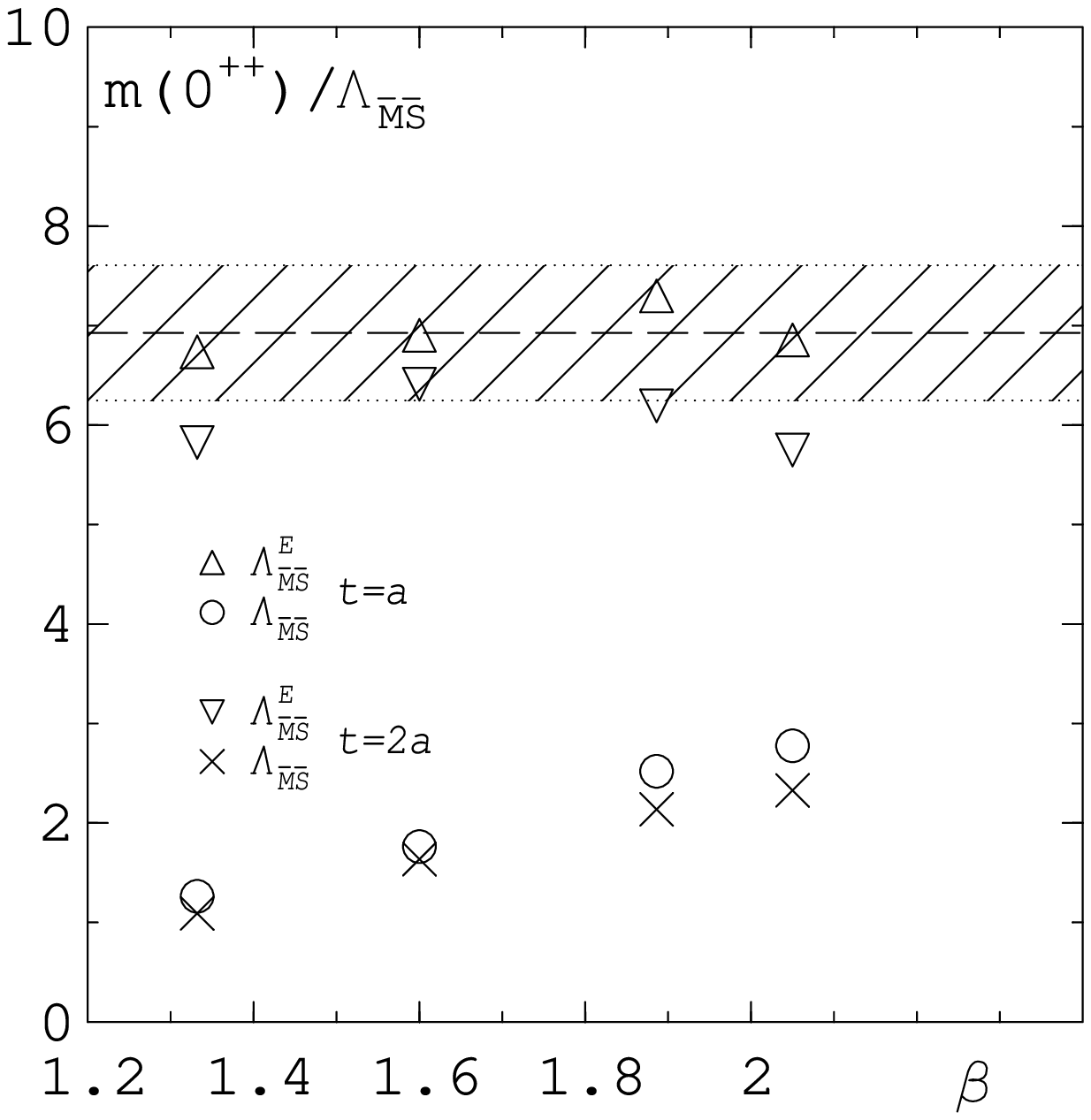}
     }
\end{center}
\vskip -3.0truecm
\caption{Asymptotic scaling of the string tension determined from
         the potential and Polyakov loops (with finite--size correction)
         and the scalar glueball mass. In the left figure the hatched
         region shows the continuum value 
         $\protect\sqrt{\sigma}/\Lambda_{\overline{\mbox{MS}}}=1.79(12)$
         determined in Ref.~\protect\cite{scaling}.
         In addition the corresponding region in the figure on the right
         is based on the continuum value
         $m(0^{++})/ \protect\sqrt{\sigma}=3.87(12)$ 
         from Ref.~\protect\cite{Mo_T}.}
\label{fig:v19}
\end{figure}

\begin{table}[htb]
\begin{center}
\begin{tabular}{|c|c|c|c|c|c|}   \hline
 $N_\tau$ & 
 $T_c/\Lambda_{\overline{\mbox{MS}}}$ &
 $\left. T_c/\Lambda_{\overline{\mbox{MS}}}\right|_E$ &
 $\sqrt{\sigma}/\Lambda_{\overline{\mbox{MS}}}$ &
 $\left. \sqrt{\sigma}/\Lambda_{\overline{\mbox{MS}}}\right|_E$ \\ \hline \hline
 2 & 0.043(1) & 0.749(20) & 0.07074(40) & 1.2363(69)\\ \hline
 4 & 0.198(2) & 1.058(11) & 0.29172(~4) & 1.5562(~6)\\ \hline
 8 & 0.380(6) & 1.102(18) & 0.56025(~8) & 1.6231(~1)\\ \hline
\end{tabular}
\end{center}
\caption{Asymptotic (2--loop) scaling of $T_c$, determined from the
         critical couplings $\beta_c(N_\tau)$, and of the string
         tension, determined from fits to the potential. Listed are
         the results with the bare and the effective, ($\beta_E$),
         coupling schemes.}
\label{tab:AS}
\medskip\noindent
\end{table}
As expected from MCRG results, we see that the approach to asymptotic
scaling for the PPM, in constrast to the SWA, is monotonous. The convergence
rate seems comparable in both models. As for the SWA we find that using
the effective coupling scheme, $\beta_E$, improves the asymptotic
scaling behavior considerably.
This shows that it takes into account
non--perturbative effects beyond the level of negative plaquettes.
Similar behavior is found for string tension and $0^{++}$
glueball mass, as shown in Fig.~\ref{fig:v19}.

\begin{figure}[htb]
\begin{center}
\vskip -1.0truecm
\leavevmode
\epsfysize=340pt\epsfbox{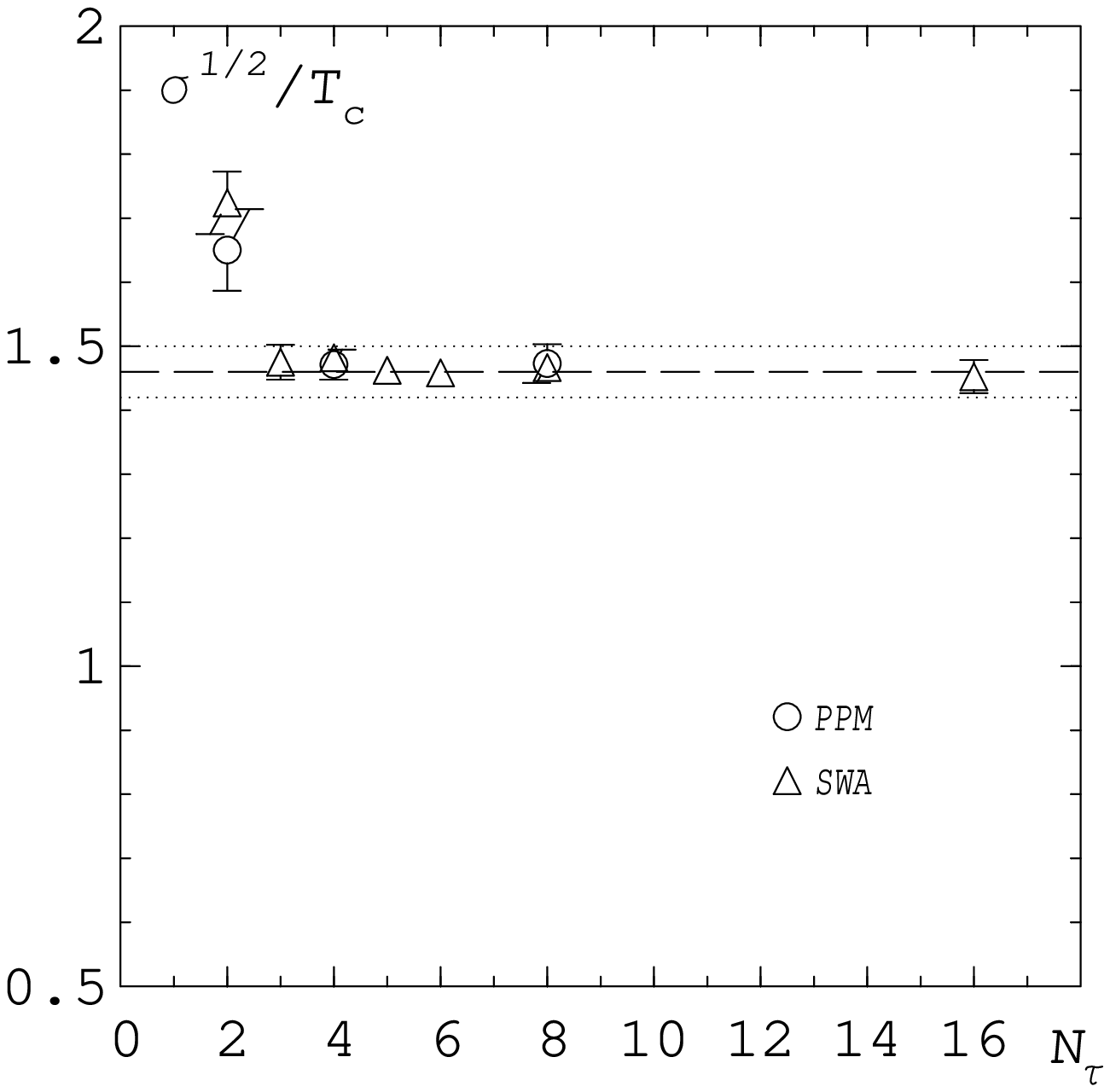}
\end{center}
\vskip -3.0truecm
\caption{Scaling of the ratio of the square root of the string tension
         and the critical temperature.
         The dashed line indicates the continuum value as determined
         in Ref.~\protect\cite{scaling}.}
\label{fig:sig_tc}
\end{figure}

While the investigation of the asymptotic scaling behavior is important to
make the connection with weak coupling perturbation theory,
it is the ``scaling'' behavior that
is important to determine the continuum limit. As usual we denote
as ``scaling'' when dimensionless ratios of physical quantities become
independent of the bare coupling --- or equivalently the lattice spacing
$a$. Having, among other physical observables, measured the deconfinement
transition temperature $T_c=1/(a(\beta_c(N_\tau))N_\tau)$ and the string
tension, we can test for scaling of the ratio
$\sqrt{\sigma}/T_c=\sqrt{\sigma}~a(\beta_c(N_\tau))N_\tau$. This
is listed in Table \ref{tab:scaling1} and shown in Fig. \ref{fig:sig_tc}.
As for the standard SU(2) lattice gauge theory, scaling is well obeyed
between the critical couplings for $N_\tau =4$ and 8. In addition, within
errors the ratio $\sqrt{\sigma}/T_c$ in the PPM is the same as for the SWA,
$\sqrt{\sigma}/T_c=1.45(4)$ \cite{scaling}. This shows that the two lattice
models will have the same continuum limit.

\begin{table}[htb]
\begin{center}
\begin{tabular}{|c|c|c|c|}   \hline
 $N_\tau$ & $\beta_c$ & $\sigma$ & $\sqrt{\sigma}/T_c$ \\ \hline \hline
 2 & 0.100(10) & 0.6810(16) & 1.65(6) \\ \hline
 4 & 1.332(~4) & 0.1353(14) & 1.48(2) \\ \hline
 8 & 1.886(~6) & 0.0339(~3) & 1.47(3) \\ \hline
\end{tabular}
\end{center}
\caption{Scaling of the ratio of the string tension from fits to
         the potential and the critical temperature.}
\label{tab:scaling1}
\medskip\noindent
\end{table}
\begin{figure}[htb]
\begin{center}
\vskip -1.0truecm
\leavevmode
\hbox{
\epsfysize=260pt\epsfbox{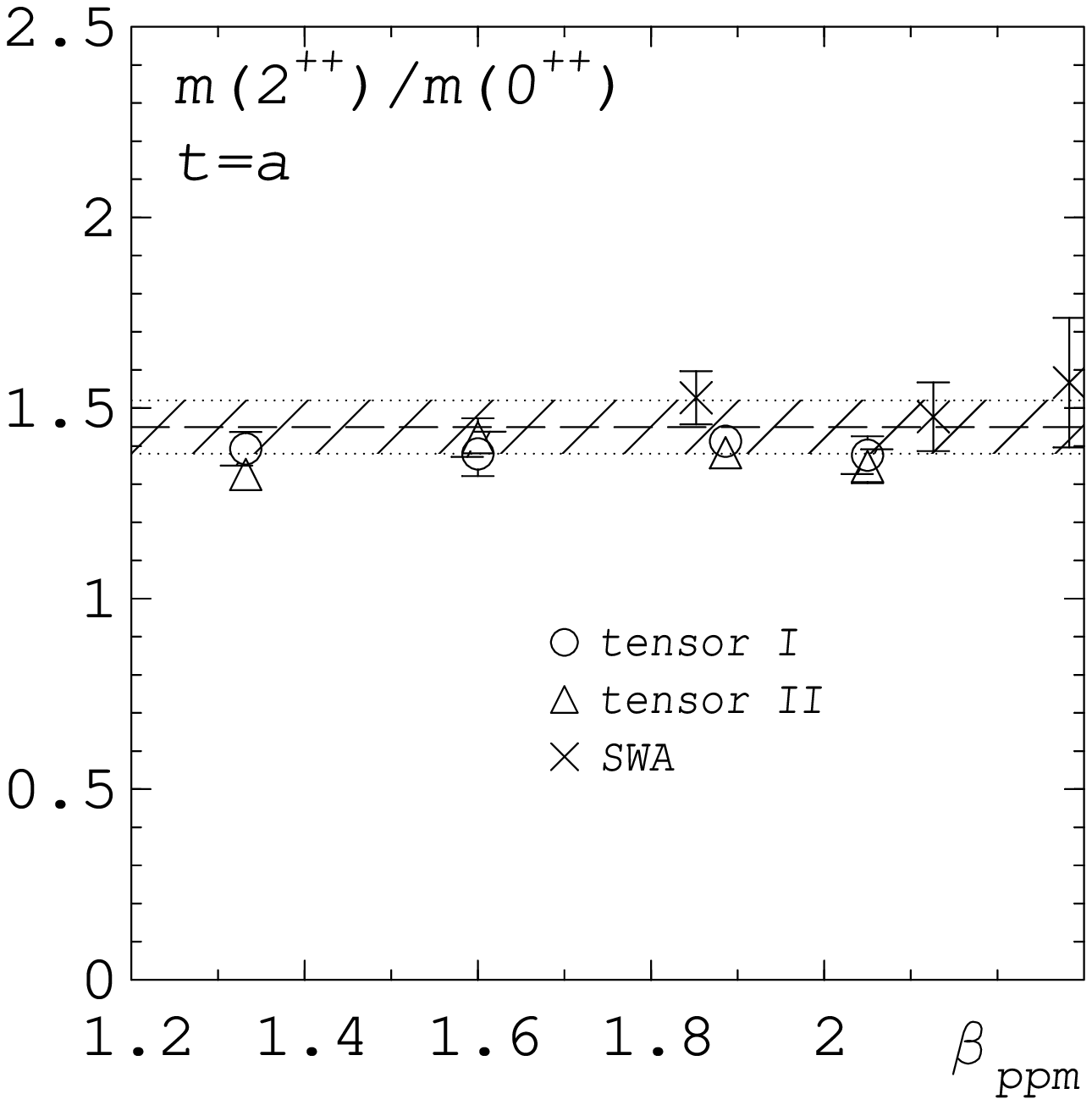}\hskip -1.0truecm
\epsfysize=260pt\epsfbox{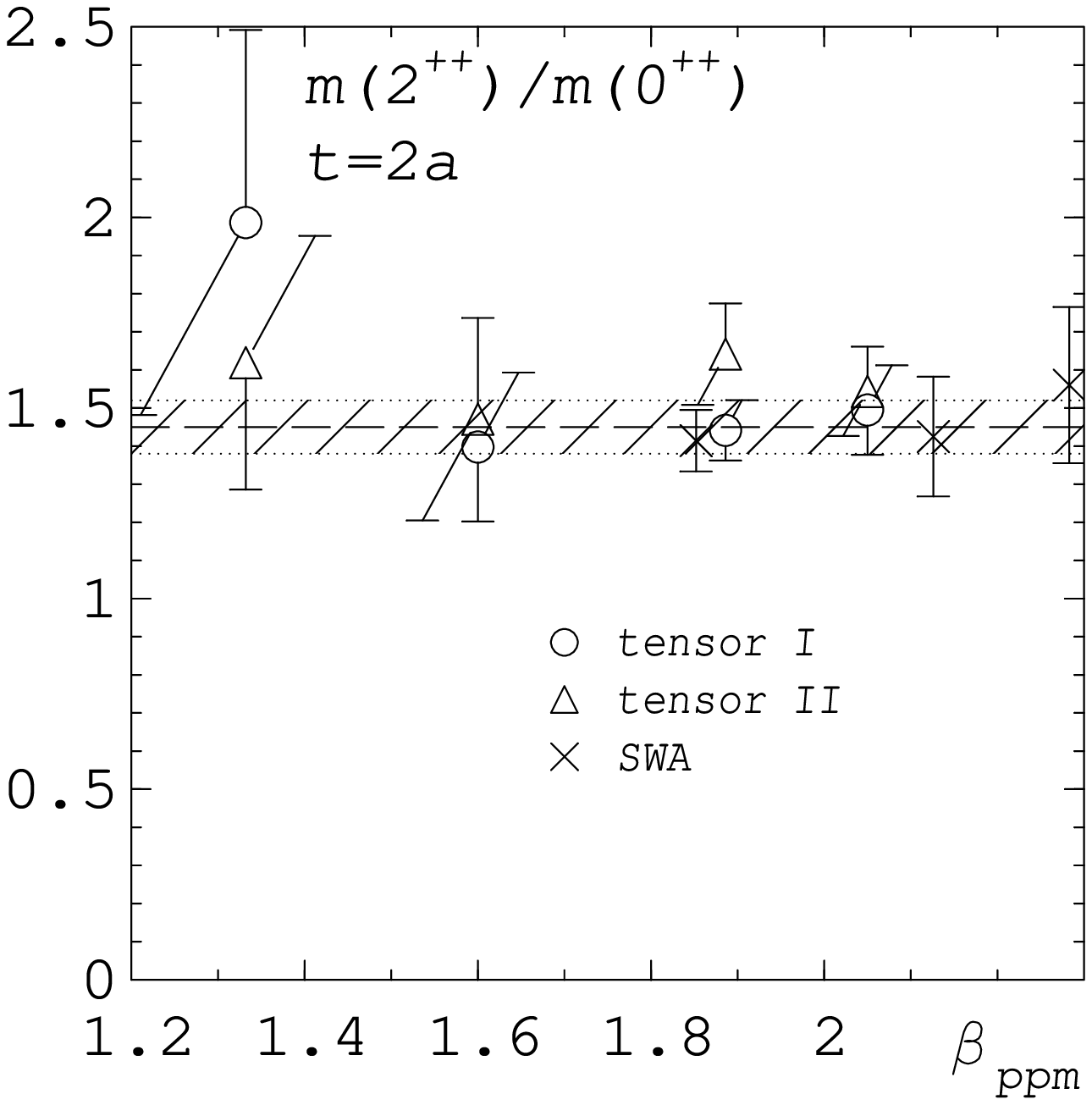}
     }
\end{center}
\vskip -2.4truecm
\caption{Scaling of the ratio of the tensor and the scalar glueball mass.
         The hatched region indicates the continuum value for the SWA
         from Ref.~\protect\cite{Mo_T}.
         Results at $\beta_{SWA}$=2.5, 2.7 and 2.9 are
         from~\protect\cite{M_T}.
         The conversion from $\beta_{SWA}$ to $\beta_{PPM}$ was done
         by matching the string tension from the heavy quark potential.}
\label{fig:v13_14}
\end{figure}
We can test for scaling using other ratios: Fig. \ref{fig:v13_14} shows the
ratio of the tensor and the scalar glueball mass. Scaling is seen, within
errors, for $\beta > 1.3$. Comparing with results from the SWA by Michael
and Teper~\cite{MiTep} we see that the PPM value is lower when the glueball
masses are extracted from $t=a$ but are fully consistent with the
continuum result $m(2^{++})/m(0^{++})=1.45(7)$ from Ref.~\cite{Mo_T}
when extracted from $t=2a$.
\begin{figure}[htb]
\begin{center}
\vskip -1.0truecm
\leavevmode
\epsfysize=340pt\epsfbox{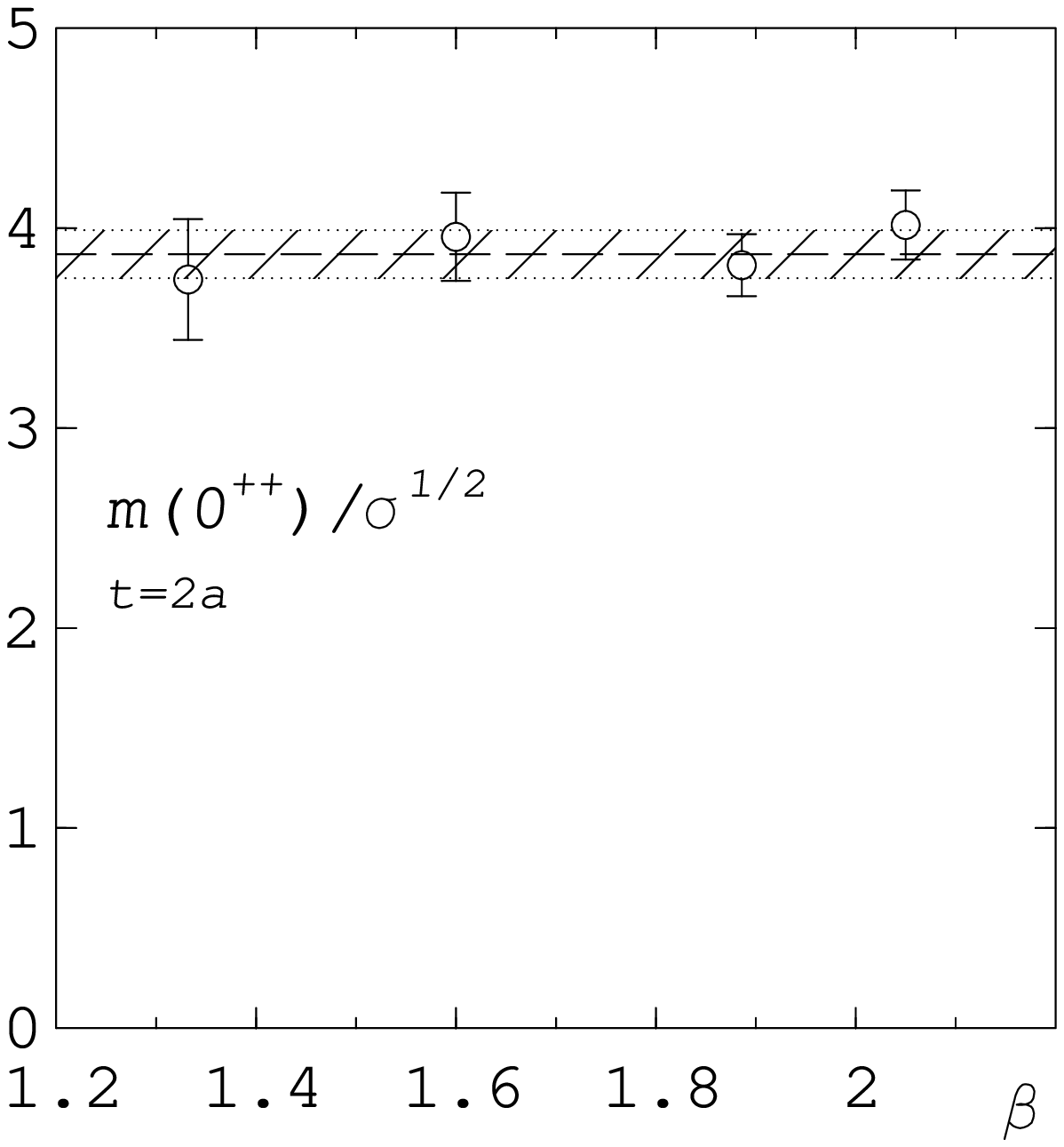}
\end{center}
\vskip -3.0truecm
\caption{Scaling of the ratio of the scalar glueball mass and
         the square root of the string tension.
         The hatched region corresponds to the continuum value
         $m(0^{++})/ \protect\sqrt{\sigma}=3.87(12)$ 
         from Ref.~\protect\cite{Mo_T}.}
\label{fig:v18}
\end{figure}
The PPM value for the ratio $m(0^{++})/\sqrt{\sigma}$, shown in
Fig.~\ref{fig:v18}, again shows scaling and looks consistent with the result
for the SWA by Michael and Teper of $3.87(12)$, as quoted in
\cite{Mo_T}.

\section{Topology}\label{sec:topology}

An important (global) property of a gauge field configuration in the
continuum is its integer valued topological charge. On the lattice the
unique assignment of an integer valued topological charge is hampered by
lattice artifacts. Indeed, for an SU(2) lattice gauge theory short distance
fluctuations on the size scale of one lattice spacing, so called dislocations,
are believed to dominate the topological charge and the 
corresponding susceptibility on ensembles of configurations produced in a
numerical simulation \cite{pt,desy}.

The occurance of negative plaquettes in configurations distributed
according the SWA is a short distance lattice artifact. The suppression of
negative plaquettes might also suppress (some of) the dislocations
afflicting the topological susceptibility measurements. To check this we
have measured the topological charge and susceptibility in our numerical
simulations of the PPM.

The dislocations manifest themselves for example by the fact that even the
geometric Philipps--Stone algorithm \cite{Ph_St}, can not assign a unique
topological charge to a given gauge field configuration. Indeed, there are
8 ways to cut hypercubes into simplices using the differently oriented
hypercube body--diagonals, and with this 8 different measurements of the
topological charge. Were the charge assignment unique all 8 calculations
would give the same result. Due to the presence of dislocations this is not
the case. To get a measure of the non--uniqueness we computed the average
covariance
\beq
CV_{PS} = \frac{1}{28} \sum_{1 \leq i < j \leq 8} \frac{ \langle \left(
Q_i - \langle Q_i \rangle \right) \left( Q_j - \langle Q_j \rangle \right)
\rangle }{ \sqrt{ \langle \left(Q_i - \langle Q_i \rangle \right)^2
\rangle \langle \left(Q_j - \langle Q_j \rangle \right)^2 \rangle }}
\label{eq:CVQ}
\eeq
between the 8 different measurements --- we used an implementation of the
Philipps--Stone algorithm \cite{PS_DESY} generously provided to us by
Micheal M\"uller-Preussker --- on $6^4$ and $8^4$ lattices at  the critical
coupling for the $N_\tau=4$ deconfinement transition. We found 0.404(15) and
0.397(13) for the SWA and 0.519(16) and 0.531(15) for the PPM, whereas a
unique charge assignment would give unity. This shows that while the PPM,
at equal lattice spacing, does somewhat better than the SWA, it still
contains dislocations.

Since some dislocations are still present we used the cooling method
\cite{cooling} to eliminate the short distance fluctuations. Already after
5 cooling sweeps the average covariance, $CV_{PS}$,
between the 8 different measurements was bigger than 0.99.

The geometric Philipps--Stone algorithm is very time consuming. Therefore we
measured, on cooled configurations, also the naive, non--integer topological
charge with a ``clover'' definition for $F_{\mu\nu}$ on the lattice.
Already after 5 cooling sweeps the covariance between naive and geometrical
topological charge definitions was about 0.96 for both PPM and SWA, and
slowly increasing with further cooling sweeps. By rounding the non--integer
charge, if the absolute value is bigger than 0.1, to the nearest integer
away from zero (``ceiling'' of the absolute value) we can assign an integer
charge to each configuration. This procedure decreases the covariance to
the geometrical charge slightly: after 5 cooling sweeps it is now about
0.91. However, out of the few variants that we tried to assign an integer
charge from the naive non--integer charge, this version gave the best
agreement of the susceptibility with that from the geometrical charge,
after 5 (and more) cooling sweeps.

In the production runs for all the other measurements, described earlier,
we measured the naive topological charge on cooled lattices, after 10, 15
and 20 cooling sweeps. We computed the topological susceptibility both with
the non--integer charge and the integer charge obtained from it by rounding
away from zero. They are listed in Table \ref{tab:Top_susc}. For the $8^4$
and $12^4$ configurations we also used the geometric Philipps--Stone
algorithm for the topological charge measurement. We did these
measurements on the ``hot'' configurations, and after 15 coolings sweeps.
The latter helps in estimating the systematic uncertainties that result
from use of a naive definition of the topological charge, even on cooled
configurations. The large drop of the susceptibility between hot and cooled
configurations indicates that on hot configurations it
is dominated by short distance dislocations, even in the PPM.

\begin{table}[htb]
\begin{center}
\begin{footnotesize}
\begin{tabular}{|l|r||l|l|l|l|l|l|l|l|}    \hline
 $\beta$ & $L$ & Hot, PS & 10c, N1 & 10c, N2 & 15c, PS & 15c, N1 &
   15c, N2 & 20c, N1 & 20c, N2 \\ \hline \hline
 1.332 &  8 & 19.3(1.0) & 2.90(13) & 5.12(20) & 4.41(20) & 2.85(13) &
   4.61(19) & 2.79(12) & 4.28(18) \\ \cline{2-10}
       & 12 & 17.5(5) & 2.90(9) & 4.00(11) & 4.51(14) & 2.89(9) &
   3.98(11) & 2.83(9) & 3.88(11) \\ \cline{2-10}
       & 16 & & 2.97(9) & 3.58(10) & & 2.91(9) & 
   3.49(10) & 2.81(9) & 3.38(10) \\ \hline
 1.6   &  8 & 5.47(19) & 1.28(5) & 2.11(7) & 1.82(7) & 1.24(5) &
   1.88(7) & 1.19(5) & 1.74(7) \\ \cline{2-10}
       & 12 & 5.44(17) & 1.36(4) & 2.08(6) & 1.92(6) & 1.35(4) &
   1.98(6) & 1.32(4) & 1.87(5) \\ \cline{2-10}
       & 16 & & 1.37(6) & 1.78(7) & & 1.39(6) & 
   1.80(7) & 1.37(6) & 1.78(7) \\ \hline
 1.8   &  8 & 1.55(7) & 0.24(2) & 0.39(2) & 0.33(2) & 0.22(1) &
   0.35(2) & 0.20(1) & 0.31(2) \\ \cline{2-10}
       & 12 & 1.20(8) & 0.60(3) & 0.91(4) & 0.80(4) & 0.61(3) &
   0.85(4) & 0.61(3) & 0.83(4) \\ \hline
 1.886 &  8 & 0.84(6) & 0.115(16) & 0.190(25) & 0.164(23) & 0.107(16) &
   0.168(23) & 0.098(15) & 0.159(23) \\ \cline{2-10}
       & 12 & 1.28(7) & 0.320(20) & 0.463(27) & 0.408(26) & 0.323(21) &
   0.437(27) & 0.323(21) & 0.423(27) \\ \cline{2-10}
       & 16 & & 0.377(17) & 0.542(22) & & 0.380(17) &
   0.517(22) & 0.318(17) & 0.505(21) \\ \hline
 2.0   &  8 & 0.41(3) & 0.031(5) & 0.051(8) & 0.047(7) & 0.028(5) &
   0.047(7) & 0.025(4) & 0.039(7) \\ \cline{2-10}
       & 12 & 0.54(3) & 0.141(13) & 0.200(19) & 0.183(17) & 0.144(14) &
   0.194(19) & 0.144(14) & 0.189(19) \\ \hline
 2.05  &  8 & 0.29(3) & 0.022(9) & 0.037(13) & 0.032(12) & 0.020(9) &
   0.032(12) & 0.017(8) & 0.027(12) \\ \cline{2-10}
       & 12 & 0.40(3) & 0.094(11) & 0.130(16) & 0.121(15) & 0.094(11) &
   0.124(16) & 0.092(11) & 0.116(14) \\ \cline{2-10}
       & 16 & & 0.157(12) & 0.215(16) & & 0.159(12) &
   0.205(16) & 0.160(12) & 0.202(15) \\ \hline
 2.2   & 12 & 0.133(10) & 0.013(4) & 0.018(5) & 0.015(4) & 0.012(4) &
   0.016(4) & 0.012(4) & 0.015(4) \\ \cline{2-10}
       & 16 & & 0.051(7) & 0.065(8) & & 0.052(7) &
   0.063(8) & 0.051(7) & 0.061(7) \\ \hline
\end{tabular}
\end{footnotesize}
\end{center}
  \caption{Topological susceptibility $\times 10^4$ and cooling. ``PS''
  denotes the   Phillips--Stone charge, ``N1'' the naive (non--integer)
  charge and ``N2''   the naive charge but rounded away from zero.}
  \label{tab:Top_susc}
\medskip\noindent
\end{table}

\begin{figure}[htb]
\begin{center}
\vskip -1.0truecm
\leavevmode
\epsfysize=360pt\epsfbox{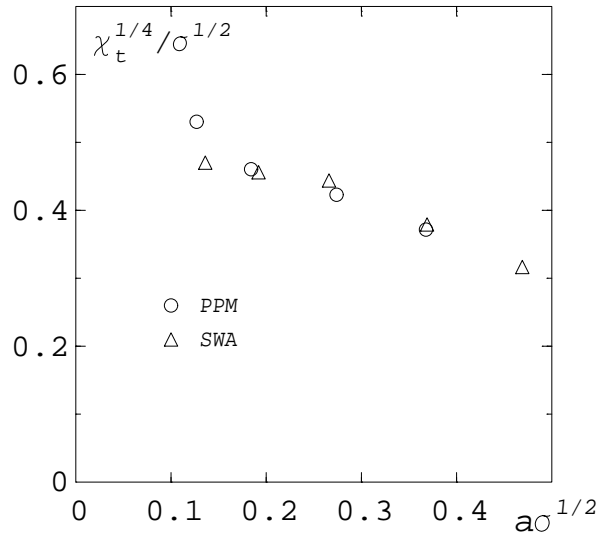}
\end{center}
\vskip -3.0truecm
\caption{Scaling test of the ratio of the topological susceptibility
         (15c,N2) and the square root of the string tension as a
         function of $a\protect\sqrt\protect\sigma$.
         Results for the SWA were obtained by Michael and Teper in
         Ref.~\protect\cite{M_T}.}
\label{fig:j4}
\end{figure}
Fig.~\ref{fig:j4} shows that the dimensionless ratio
$R_t =\chi_t^{1/4}/\sqrt{\sigma}$ is a smooth
function of the lattice spacing consistent with a continuum limit
$R_t \approx 0.5$. We used the naive charge rounded away from zero (N2)
after 15 cooling steps on the $16^4$ lattices. This definition gave
consistent results with its counterpart on the $12^4$ lattices and the
topological charge measured with the geometric Philipps--Stone algorithm.

A comparison with results from Ref.~\cite{M_T} shows
that the results for the PPM and those obtained with the SWA agree very well.
Only the PPM value of $R_t$ at the largest coupling $\beta=2.05$,
corresponding to the smallest $a \sqrt \sigma$,
seems to
be too high. The reason for this is again that the lattice size, $L=16$,
might be too small.

\section{Conclusion}\label{sec:conclusion}

We made a comprehensive study of scaling properties at
zero and finite temperatures in the pure gauge $SU(2)$ theory 
with negative plaquettes suppressed (PPM), 
aimed at the comparison with the corresponding properties
in the standard Wilson theory.
This study included the calculation of the heavy quark potential,
the masses of $0^{++}$ and $2^{++}$ glueballs, the ``step
$\beta$--function'', $\Delta \beta(\beta)$, and the topological charge
with its susceptibility.
To monitor the finite temperature phase transition we calculated
the Polyakov loops with  their susceptibility and Binder cumulant.
The calculations were done on symmetric lattices 
with size up to $16^4$ and on asymmetric ones with
temporal extent $N_{\tau} = 2, \, 4$ and $8$.
The  values of the gauge coupling  $\beta$ we used covered
the interval between zero and $2.6$.

Our main conclusion is that the PPM belongs to the same 
universality class as the standard Wilson theory.
For $\beta < \beta^*(L)$, where $\beta^*(16) \sim 2.05$,
the heavy quark potential, $V(R)$, exhibits the
linearly rising, confining behavior at large enough $R$.
The physical ratios $m(2^{++}):m(0^{++}):\sqrt{\sigma}:\chi_t^{1/4}$
(the latter computed after cooling) show similar scaling patterns
in the PPM and the conventional formulation.

In the finite temperature sector of the PPM we find a clear signal for
a $2^{\rm nd}$ order phase transition with critical exponents consistent
with the 3--dimensional Ising values. This shows that the PPM modification
to the SWA maintains the universality class, a necessary condition
to assure that the PPM and the theory with SWA describe the same 
physics in the continuum limit.

However, the approach to the asymptotic scaling behavior changed
drastically after the suppression of one type of lattice artifacts,
namely the occurance of negative plaquettes in the gauge field
configurations. In contrast to the SWA case the dependence of
$T_c/\Lambda_{\overline{\mbox{MS}}}$ and
$\sqrt{\sigma}/\Lambda_{\overline{\mbox{MS}}}$
on the lattice coupling or, equivalently, the lattice spacing is monotonous. 
Furthermore the dip of the SWA step $\beta$--function is completely removed.
The PPM step $\beta$--function comes steeply from above and converges rather
slowly to the perturbative 2--loop value. This shows that phenomena
connected to the occurance of negative plaquettes are the origin of
this structure. The effective coupling
scheme derived from the plaquette expectation value
improves asymptotic scaling equally well for the PPM and the SWA.

As was expected, for the computation of the topological charge with the
geometric Philipps--Stone algorithm on ``hot'' configurations,
the modified action does somewhat better, compared to the
Wilson action, because the configurations are somewhat smoother.
Nevertheless, the suppression of negative plaquettes               
does not resolve the problem of rough configurations completely.
Despite the fact that it removed those dislocations discussed
in~\cite{pt,desy} that have been shown to cause the topological
susceptibility, $\chi_t$, to diverge in the
continuum limit, the presence of other dislocations leaves
the fate of the continuum limit of $\chi_t$ an open question.
The important point is that lattice artifacts can exist 
at {\it other} scales independently of the small size 
artifacts, like the occurance of negative plaquettes (at least,
in nonabelian theories). Presumably, further modifications of the
gauge action are necessary to control all dislocations.

The work of JF and UMH was supported in part by the DOE under grants
\#~DE-FG05-85ER250000 and \#~DE-FG05-92ER40742,
and the work of VKM was supported by the Deutsche
Forschungsgemeinschaft under research grant Mu 932/1-2.
The computations have been
carried on the cluster of Alpha workstations at SCRI and on
the Convex at Humboldt--Universit\"at zu Berlin. 
The simulations to determine the $N_\tau = 8$ critical
coupling were done on the CM-2 at SCRI.

\section{Appendix}\label{sec:appendix}

It has been observed that the bare lattice coupling is a ``bad'' coupling
constant to investigate asymptotic scaling. Most alternative couplings are
extracted from the measured average plaquette values, which we list in
Table \ref{tab:Plaq}.

\begin{table}[htb]
\begin{center}
\begin{tabular}{|l||l|l|l|}    \hline
 $\beta$ & $L=8$ & $L=12$ & $L=16$ \\ \hline \hline
 0.1   & 0.54706(4) & & \\ \hline
 1.332 & 0.41695(7) & 0.41685(2) & 0.41685(1) \\ \hline
 1.6   & 0.38753(7) & 0.38756(2) & 0.38755(2) \\ \hline
 1.8   & 0.36651(7) & 0.36666(3) & \\ \hline
 1.886 & 0.35773(7) & 0.35785(3) & 0.35792(2) \\ \hline
 2.0   & 0.34651(6) & 0.34656(3) & \\ \hline
 2.05  & 0.34161(6) & 0.34173(3) & 0.34173(2) \\ \hline
 2.2   & 0.32746(6) & 0.32740(3) & 0.32747(2) \\ \hline
 2.4   & 0.30912(4) & 0.30913(3) & \\ \hline
 2.6   & 0.29182(4) & 0.29182(3) & 0.29181(1) \\ \hline
\end{tabular}
\end{center}
\caption{Average plaquette values, $\langle 1-\Tr U_p/2 \rangle$.}
\label{tab:Plaq}
\medskip\noindent
\end{table}

Parisi was the first to suggest the use of an effective 
coupling, $\beta_{eff}$, instead of the bare coupling $\beta$ \cite{Beta_E}.
In our analysis we used the effective coupling $\beta_E$, obtained
from the plaquette by \cite{K_P}
\beq
1/\beta_E = \frac{4}{3} \langle 1-\Tr U_p/2 \rangle ~~~.
\eeq
More recently, for a systematic method of improving the lattice
perturbation theory, Lepage and Mackenzie advocated to extract a coupling
$g^2_V$ --- for SU(2) we have the relation $\beta_X = 4 / g^2_X$ for a
scheme $X$ --- to be interpreted as a running coupling at a scale $3.41/a$,
from \cite{L_M}
\beq
-\log \langle \Tr U_p/2 \rangle = \frac{3}{16} g^2_V ( 1 - 0.0629 g^2_V ) .
\eeq
The two different couplings are tabulated in Table \ref{tab:Eff_coup}. In
perturbation theory the two different schemes are related by
\beq
\beta_E = \beta_V + 0.6264 + {\cal O} (1/\beta_V) .
\eeq
We see that the actual values are not too far off from this relation and
certainly compatible with the ${\cal O} (1/\beta_V)$ correction. It is
interesting to note that under the same physical conditions, the deconfinement
transition for given $N_\tau$, the improved couplings of the PPM are quite
close to their counterparts with SWA.

\begin{table}[htb]
\begin{center}
\begin{tabular}{|l||l|l|l|l|}    \hline
 $\beta$ & $\beta_E$ & $g^2_E$ & $\beta_V$ & $g^2_V$ \\ \hline \hline
 0.100 & 1.37096 & 2.91765 & & \\ \hline
 1.332 & 1.79921 & 2.22320 & 1.06091 & 3.77036 \\ \hline
 1.600 & 1.93518 & 2.06699 & 1.21222 & 3.29973 \\ \hline
 1.800 & 2.04549 & 1.95552 & 1.33190 & 3.00323 \\ \hline
 1.886 & 2.09585 & 1.90853 & 1.38590 & 2.88621 \\ \hline
 2.000 & 2.16413 & 1.84832 & 1.45862 & 2.74232 \\ \hline
 2.050 & 2.19472 & 1.82256 & 1.49103 & 2.68270 \\ \hline
 2.200 & 2.29078 & 1.74613 & 1.59229 & 2.51211 \\ \hline
 2.400 & 2.42616 & 1.64869 & 1.73384 & 2.30701 \\ \hline
 2.600 & 2.57008 & 1.55637 & 1.88320 & 2.12404 \\ \hline
\end{tabular}
\end{center}
\caption{Effective couplings from the $L=12$ plaquette values, except for
         $\beta =0.1$ where the $L=8$ value was used.}
\label{tab:Eff_coup}
\medskip\noindent
\end{table}

\end{document}